\documentclass[fleqn,usenatbib]{mnras}

\usepackage{newtxtext,newtxmath}

\usepackage[T1]{fontenc}

\DeclareRobustCommand{\VAN}[3]{#2}
\let\VANthebibliography\thebibliography
\def\thebibliography{\DeclareRobustCommand{\VAN}[3]{##3}\VANthebibliography}

\usepackage{graphicx}	
\usepackage{lscape}
\usepackage{longtable}
\usepackage{bm}

\newcommand{\kms}{$\rm km\,s^{-1}$}

\title[Spatial orientation and shape of the velocity ellipsoids]{Spatial orientation and shape of the velocity ellipsoids of the Gaia DR3 giants and subgiants in the Galactic plane}

\author[Dmytrenko et al.] {
A. M. Dmytrenko,$^{1}$\thanks{E-mail: astronom.karazin007@gmail.com (AMD)}
P. N. Fedorov,$^{1}$\thanks{E-mail: pnfedorov@gmail.com (PNF)}
V. S. Akhmetov,$^{1,2}$\thanks{E-mail: akhmetovvs@gmail.com (VSA)}
A. B. Velichko,$^{1}$
S. I. Denyshchenko,$^{1}$
\newauthor
V. P. Khramtsov,$^{1}$
I. B. Vavilova,$^{3}$
D. V. Dobrycheva,$^{3}$
O. M. Sergijenko,$^{3}$
A. A. Vasylenko$^{3}$
 \newauthor
and O.V. Kompaniiets$^{3}$
\\
$^{1}$Institute of astronomy of V.N.Karazin Kharkiv national university, Svobody sq. 4, Kharkiv, 61022, Ukraine \\
$^{2}$INAF-Osservatorio Astrofisico di Torino, Via Osservatorio 20, Pino Torinese, Turin, I-10025, Italy \\
$^{3}$Main Astronomical Observatory of the National Academy of Sciences of Ukraine, Akademik Zabolotnyi 27, Kyiv, 03143, Ukraine
}

\date{Accepted XXX. Received YYY; in original form ZZZ}

\pubyear{2025}

\begin{document}
\label{firstpage}
\pagerange{\pageref{firstpage}--\pageref{lastpage}}
\maketitle

\begin{abstract}
We present the results of determining the parameters characterizing the shape and orientation of residual velocity ellipsoids from the \textit{Gaia} DR3 red giants and subgiants. We show the distribution of velocity dispersions $\sigma^2_\xi, \sigma^2_\eta, \sigma^2_\zeta$ in the Galactic plane obtained from three components of the spatial velocity, as well as the coordinate distribution of the intersection points of the velocity ellipsoid axes with the celestial sphere, in particular the deviations of the longitudes and latitudes of the vertices of stellar regions located within spheres with a radius of 1 kpc centered in the Galactic mid-plane. The area of the Galactic disk under study is in the range of Galactocentric coordinates $0<R<15$ kpc and $120^\circ<\theta<240^\circ$. We show that the vertex deviations in some regions of the Galactic mid-plane can reach $30^\circ$ in longitude, and $15^\circ$ in latitude. This indicates the presence of kinematic distortions of the stellar velocity field, especially noticeable in the angular range of $150^\circ<\theta<210^\circ$ at a distance of approximately 13 kpc. We propose the angles of deviation of longitudes and latitudes of the ellipsoid axes of residual stellar velocities to be considered as kinematic signatures of various Galactic deformations determined from real fields of spatial velocities. We present the distribution of parameters characterizing the shapes of velocity ellipsoids, as well as their distribution of the semi-axes length ratios. We note a local feature in this distribution and in the distribution of the elongation measurements of the ellipsoids. We perform a comparison of the results obtained from the tensor of deformation velocities and from the observed spatial velocities.

\end{abstract}


\section{Introduction}
\label{sec:introduction}
The third release of the \textit{Gaia} mission -- \textit{Gaia} DR3 catalogue \citep[][]{Prusti2016, Vallenari2023} has made available new data to study the stellar kinematics not only within the solar neighborhood, but also in a vast part of the Milky Way. The presence of parallaxes, radial velocities and proper motions of stars in this release allows us to obtain new information about the stellar kinematics. The use of the \textit{Gaia}  data have increased the level of detail in the velocity distribution of stars \citep[see e.g.,][]{Antoja2018, Helmi2018}, resulting in the detection of many previously unknown features \citep[][]{Antoja2021, Drimmel2023}. The availability of parallaxes and radial velocities, together with positions and proper motions of stars, allows us to analyze the three-dimensional velocity field  $\bm{V}_{\rm OBS}(\bm{r})$, turned out to be especially valuable for new kinematic studies. This has given rise to new opportunities for determining some important kinematic parameters. For example, studies of the shapes of residual velocity ellipsoids, deformations of the Galactic disk, and coordinates of the kinematic centers of rotation of various stellar systems, which are closely related to the symmetry and shape of the Galactic potential \citep[][]{Amendt1991, Kuijken1991, Smith2012} can now provide huge amount of material for galactic dynamics.

Although there are certain difficulties in interpreting the obtained results, caused, for example, by the discrepancy between the distances to the sources determined from the \textit{Gaia} parallaxes and using the Bayesian method \citep[][]{Bailer-Jones2021}, the use of different estimates of distances to the Galactic center $R_0$ or the relatively small number of stars with known radial velocities ($\sim$33 million),  the solutions of the kinematic problems mentioned above are beyond doubt.

In this paper, we focus on determining the parameters of the velocity ellipsoids whose centroids are located in the Galactic mid-plane. In order to determine the ellipsoid orientations (directions of their axes) and the ellipsoid shapes (lengths of their semi-axes), we study the three-dimensional distribution of residual velocities of the \textit{Gaia} DR3 red giants and subgiants. This study is an extension of our previous work \citep[][]{Dmytrenko2023}, based on the analysis of the deformation velocity tensor components. Because in stellar dynamics the components of the deformation velocity tensor are related to the forces causing these deformations, \citet{Dmytrenko2023} proceeded from the assumption that the orientation of the principal axes of the deformation velocity tensor is determined by the directions of the forces causing the deformation movements. At the same time, it should be kept in mind that the deformation velocity is only one of the components of the full velocity field, which can cause such observable kinematic phenomena as, for example, vertex deviation. Therefore, in this work, we analyze velocity ellipsoids which are determined by the stellar spatial velocities, and compare the obtained results with those obtained based on the analysis of the deformation velocity tensor.

This paper is organized as follows. In Section \ref{sec:cs}, we review the coordinate systems used in the paper and describe the sample of stars from which the centroids and velocity ellipsoids have been formed. In Section \ref{sec:method}, we present the method used to obtain the parameters of the velocity ellipsoids and their uncertainties. Section \ref{sec:results} describes and discusses the results obtained. In addition, we provide the comparison of the orientation parameters of the velocity ellipsoids with similar parameters obtained from the velocity deformation tensors (Section \ref{sec:comparison}). In Section \ref{sec:conclusions} we briefly summarize the results obtained.

\section{Coordinate systems}
\label{sec:cs}

\begin{figure}
    \centering
    \includegraphics[width=1.0\linewidth]{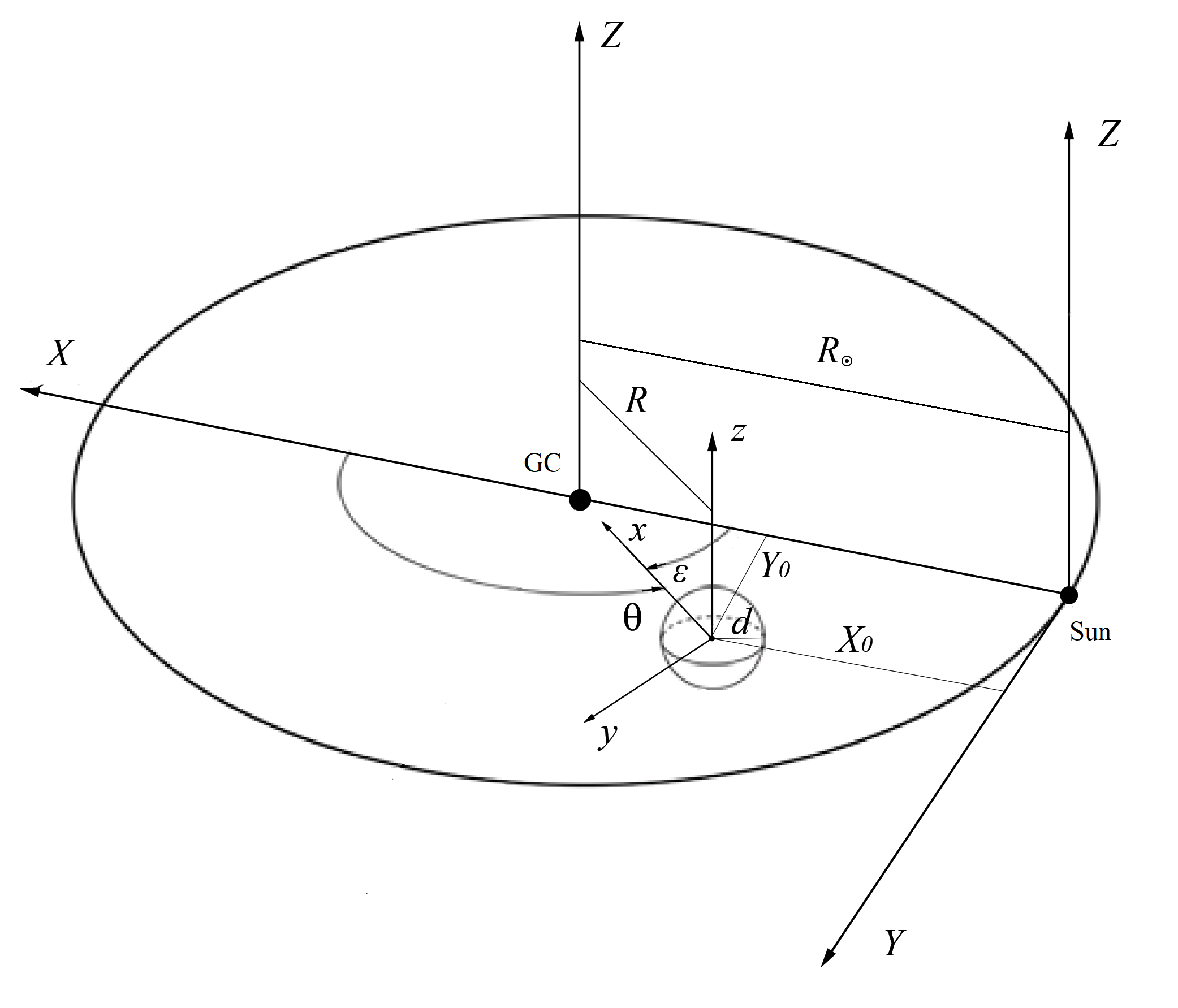}
    \caption{The link between the Galactocentric, Galactic and local coordinate systems.}
    \label{fig:cs}
\end{figure}

As noted by \citet{Fedorov2021, Fedorov2023}, the local coordinate system similar to the rectangular Galactic system $XYZ$ with the origin at the barycenter of the Solar System can be introduced at any arbitrary point of the Galactic mid-plane (see Fig. \ref{fig:cs}), provided that the spatial coordinates and components of the spatial velocity are known for this point (for example, a star) and for the stars in its vicinity. The transition from the Galactic Cartesian coordinate system to the local Cartesian system is equivalent to moving a fictitious observer from the barycenter of the Solar System to the origin of the local system, specified by its Galactic coordinates \citep{Akhmetov2024}. The local Cartesian coordinate system, as well as the Galactic Cartesian system, are right-handed orthogonal coordinate system. Their $x$-axes are always directed from a specific centroid to the center of the Milky Way, the $y$-axes coincide with the direction of the Galactic rotation, and the $z$-axes are always parallel to the direction to the north Galactic pole. The points from which a fictitious observer supposedly performs the observations are specified as follows. First, we identify spherical regions of a 1 kpc radius, the centers of which are located at the nodes of a rectangular grid coinciding with the Galactic mid-plane. Coincidence with the Galactic mid-plane is ensured by setting the condition $Z=0$ for any node. The nodes are spaced from each other along the coordinates $X$ and $Y$ at a distance of 100 pc. Around each node, we form stellar systems containing stars that fell into a sphere of 1 kpc radius, the centroids of which have velocities equal to the average velocity of the stars making up the sphere. 

To analyze the stellar velocity field, we used a sample of red giants and subgiants from \textit{Gaia} DR3, for which the spatial coordinates and components of the spatial velocity are known. For ensure these conditions, we select 33 million stars from \textit{Gaia} DR3, for which, in addition to the five main astrometric parameters, radial velocities have been measured. The distribution of the number of stars in this sample by heliocentric distance shows that the range of distances corresponding to the main-sequence stars is approximately contained in a relatively narrow interval of 7 kpc $<R<$ 9 kpc. All other stars in this sample are mainly represented by red giants. The corresponding range of Galactocentric distances is wider and lies in the interval 0 kpc $<R<$15 kpc, which provides significantly greater coverage of the Galactic plane. As in our previous work by \citet{Fedorov2023}, we exclude from this sample those stars for which the following conditions are met \citep[][]{Lindegren2018}:

\begin{equation*}
 \begin{cases}
   RUWE > 1.4,\\
   \varpi/\epsilon(\varpi) < 5,\\
   (\mu_\alpha/\epsilon({\mu_\alpha}))^2 + (\mu_\delta/\epsilon({\mu_\delta}))^2 < 25.
 \end{cases}
\end{equation*}

Additionally, using the parallax bias Z5 proposed by \citet{Lindegren2021} we correct the stellar parallaxes in our sample. In the magnitude range $M_G$ 9--13, we also correct proper motions of stars according to the recommendations from \citet{Cantat-Gaudin2021}.

Cutting off the main sequence in the Hertzsprung-Russell diagram $M_G-(BP-RP)$ with two linear functions from about 30 million stars remaining after applying the Lindegren criteria, we form a final sample of 14.2 million giants and subgiants we further use in this work. This stellar sample coincides with the sample used in the work \citet{Dmytrenko2023}, which makes it possible to compare the obtained results.

\section{Method}
\label{sec:method}
In the era before \textit{Gaia}, measuring the precise orientation of the velocity ellipsoid was difficult, mainly due to the lack of reliable parallaxes, radial velocities, and proper motions of stars outside the Solar neighborhood. Using \textit{Gaia} DR3 data, we can determine both the shape and orientation of the velocity ellipsoid fairly accurately not only in the Solar neighborhood, but also in the part of the Galaxy for which Gaia DR3 has astrometric parameters. We used our sample of stars without dividing them into thin and thick disks and determine the shape and orientation of the velocity ellipsoids using the method described below.

It is known that stellar disks are essentially anisotropic and local properties of stars, such as average velocities, are completely determined by the velocity dispersion tensor \citep{Vorobyov2008}. The Schwarzschild ellipsoidal law of velocity distribution (or simply: velocity ellipsoid) is understood as a three-dimensional Gaussian distribution of residual velocities of stars located in a certain volume.

The residual velocity of the $k$-th star ($k = 1..N$) $\bm V'_{k}=(V'_{x,k}; V'_{y,k}; V'_{z,k})$ is the velocity of the star with respect to the velocity of its centroid $\bm{V_0}$:

\begin{equation}
\bm{V'_k} = \bm{V_k} - \bm{V_0}.
\label{eq:v_residual}
\end{equation}

Obviosly, 
\begin{equation}
{\frac{1}{N} \sum\limits_{k=1}^N \bm{V'_k}=0}.
\label{eq:v_residual_sum}
\end{equation}

Let us denote the second-order moments of the residual stellar velocities (covariance and dispersion) as follows:
\begin{equation}
\begin{matrix}
\sigma_{12} = \sigma_{21} = \sigma_{x}\sigma_{y} = \frac{1}{N}\sum\limits_{k=1}^N V'_{x,k} V'_{y,k}, &
\sigma_{11} = \sigma^2_{x} = \frac{1}{N}\sum\limits_{k=1}^N (V'_{x,k})^2, \\
\sigma_{23} = \sigma_{32} = \sigma_{y}\sigma_{z} = \frac{1}{N}\sum\limits_{k=1}^N V'_{y,k} V'_{z,k}, &
\sigma_{22} = \sigma^2_{y} = \frac{1}{N}\sum\limits_{k=1}^N (V'_{y,k})^2, \\
\sigma_{13} = \sigma_{31} = \sigma_{x}\sigma_{z} = \frac{1}{N}\sum\limits_{k=1}^N V'_{x,k} V'_{z,k}, &
\sigma_{33} = \sigma^2_{z} = \frac{1}{N}\sum\limits_{k=1}^N (V'_{z,k})^2.
\end{matrix}
\label{eq:sigmas}
\end{equation}
where $V'_x$, $V'_y$, $V'_z$ are the projections of the residual velocity onto the axes of the local coordinate system $x, y, z$. 

The covariances and variances form a real symmetric tensor $\bm\Sigma$, the matrix of which has the form:
\begin{equation}
\bm\Sigma =
\begin{pmatrix}
\sigma_{11} & \sigma_{12} & \sigma_{13} \\
\sigma_{21} & \sigma_{22} & \sigma_{23} \\
\sigma_{31} & \sigma_{32} & \sigma_{33}
\end{pmatrix}.
\label{eq:matrixSigma}
\end{equation}
Each symmetric real tensor can be associated with its tensor surface. Because $\sigma_{ij}$ is always positive, such a surface will be an ellipsoid, the equation of which in rectangular coordinates is as follows:
\begin{equation}
2\sigma_{12}xy + 2\sigma_{23}yz + 2\sigma_{13}xz + \sigma_{11}x^2 + \sigma_{22}y^2 + \sigma_{33}z^2 = 1.
\label{eq:elipsoid}
\end{equation}

According to continuum mechanics \citep[][]{Tarapov2002}, we can determine the orientation of the principal axes of a tensor surface with respect to a given coordinate system. For example, the Galactic coordinates of points on the celestial sphere to which the principal axes of a symmetric tensor $\bm\Sigma$ are directed. From the linear algebra course it is known that there is such a number $\lambda$ and a vector $\\bm{x}\neq0$, meeting the following relation: 
\begin{equation} \bm\Sigma \bm{x} = \lambda \bm{x}. \label{eq:lambda_1} \end{equation} 
Here $\lambda$ is an eigenvalue $\bm\Sigma$, while $\bm{x}$ is an eigenvector, corresponding to the eigenvalue of $\lambda$. The vector equation \ref{eq:lambda_1} will be equivalent to the following matrix equation:
\begin{equation}
(\bm\Sigma - \lambda E) X = 0, X \neq 0.
\label{eq:lambda_2}
\end{equation} 
where $\bm E$ is a unit matrix, while the matrix columns $\bm X$ are eigenvectors $\bm \Sigma$. From this it follows that the number $\lambda$ will be the eigenvalue of the matrix $\bm\Sigma$ only if the determinant $(\bm\Sigma - \lambda \bm E)= 0$. That is, when $\lambda$ is a root of a polynomial $f(\lambda) = det(\bm\Sigma - \lambda \bm E)$, which is called the characteristic equation.

For a three-dimensional symmetric matrix $\bm \Sigma$ the characteristic equation $p(\lambda)$ will look as follows:
\begin{equation}
f(\lambda) =
\begin{pmatrix}
\sigma_{11}-\lambda & \sigma_{12} & \sigma_{13} \\
\sigma_{12} & \sigma_{22}-\lambda & \sigma_{23} \\
\sigma_{13} & \sigma_{23} & \sigma_{33}-\lambda
\end{pmatrix} = 0,
\label{eq:polinom_1}
\end{equation}
or
\begin{equation}
f(\lambda) = \lambda^3 + a \lambda^2 + b \lambda + c = 0,
\label{eq:polinom_2}
\end{equation}
where
\begin{subequations}
\begin{align}
a&=-(\sigma_{11} + \sigma_{22} + \sigma_{33}), \\
b&=-(\sigma_{12}^2 + \sigma_{23}^2 + \sigma_{13}^2 - \sigma_{11} \sigma_{22} - \sigma_{22} \sigma_{33} - \sigma_{11} \sigma_{ 33}), \\
c&= -(\sigma_{11} \sigma_{22} \sigma_{33} + 2 \sigma_{12} \sigma_{23} \sigma_{13} - \sigma_{12}^2 \sigma_{33} - \sigma_{23}^2 \sigma_{11} - \sigma_{13}^2 \sigma_{22}).
\end{align}
\label{eq:polinom_3}
\end{subequations}

The equation roots (\ref{eq:polinom_2}) can be calculated with the use both numerical methods and Vieta's trigonometric formula:
\begin{subequations}
\begin{align}
\lambda_1 & = 2 \sqrt{-Q} \cos \phi - \frac{a}{3}, \\
\lambda_2 & = 2 \sqrt{-Q} \cos(\phi - \frac{2\pi}{3}) - \frac{a}{3}, \\
\lambda_3 & = 2 \sqrt{-Q} \cos(\phi + \frac{2\pi}{3}) - \frac{a}{3}.
\end{align}
\label{eq:Vieta3D}
\end{subequations}
Here:
\begin{equation}
Q = \frac{3b - a^2}{9},
\phi = \frac{1}{3} \arccos \frac{R}{\sqrt{-Q^3}},
R = \frac{1}{54}(-2a^3 + 9ab -27c).
\label{eq:Vieta3D_2}
\end{equation}

For real symmetric matrices, the eigenvalues (roots of the equation) will also be real. Using the eigenvalues $\lambda_{1,2,3}=\lambda_{i}$, one can find the corresponding orthogonal eigenvectors, which form the so-called eigenbasis of the tensor. The tensor $\bm\Sigma$ in the system of its principal axes takes a diagonal form:
\begin{equation}
{\bm \Sigma}_{\rm diag} =
\begin{pmatrix}
\lambda_1&0&0\\ 0 & \lambda_2 & 0 \\ 0 & 0 & \lambda_3
\end{pmatrix},
\label{eq:lambda_3}
\end{equation}

\begin{figure}
    \centering
    \includegraphics[width=1\linewidth]{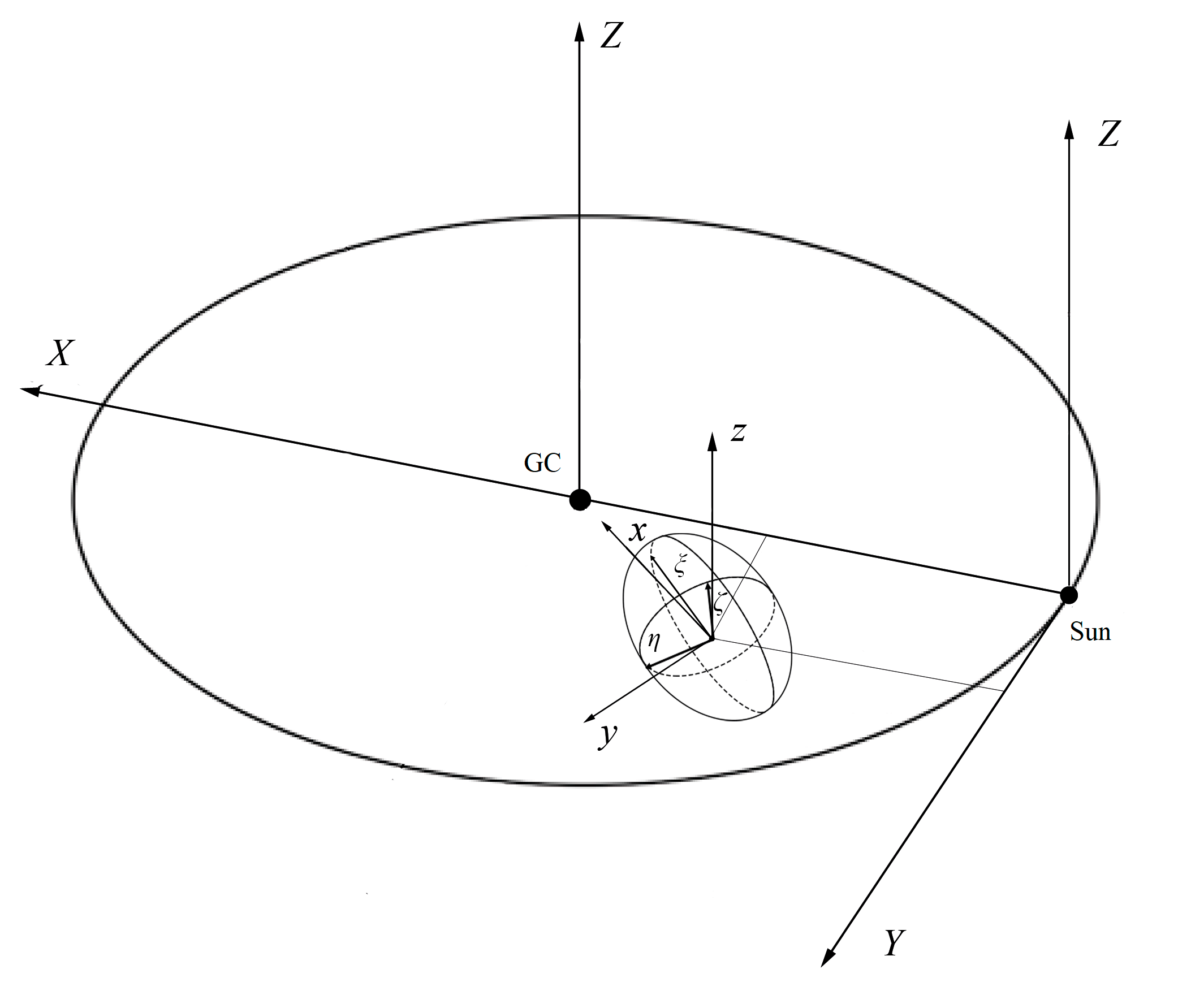}
    \caption{The coordinate systems used in the work are: Galactic $X,Y,Z$, local Galactic $xyz$ and principal axes of the ellipsoid $\xi, \eta, \zeta$.}
    \label{fig:elips_cs}
\end{figure}

If $\lambda_1, \lambda_2, \lambda_3$ are positive, such an imaginary surface is an ellipsoid with axes $\xi, \eta$ and $\zeta$ (see Fig. \ref{fig:elips_cs}), and the values $\lambda_1, \lambda_2, \lambda_3$  are the principal (eigen) values equal to the squares of the semi-axes $p, q, r$ of the residual velocity ellipsoid:
\begin{align}
   & \lambda_1 = \sigma^2_\xi = p^2,~~~\lambda_2 = \sigma^2_\eta = q^2, ~~~\lambda_3 = \sigma^2_\zeta = r^2 \nonumber \\
   & \lambda_1 > \lambda_2 > \lambda_3 \nonumber
\end{align}

As has been shown by \citet{Parenago1951}, the directions ${L_{i},B_{i}}$ of the principal axes of the ellipsoid can be found, for example, by the following formulas:
\begin{subequations}
\begin{align}
\tan L_i &= \frac{\sigma_{13}\sigma_{23}-(\sigma_{33}-\lambda_i)\sigma_{12}}{(\sigma_{22}-\lambda_i)(\sigma_{33}-\lambda_i )-\sigma^2_{23}}, \\
\tan B_i &= \frac{(\sigma_{22}-\lambda_i)\sigma_{13} - \sigma_{12}\sigma_{23}}{\sigma^2_{23}-(\sigma_{22}-\lambda_i)(\sigma_{33}-\lambda_i)}\cos L_i,
\end{align}
\label{eq:LBs}
\end{subequations}
and their uncertainties are:
\begin{subequations}
\begin{align}
\epsilon(L_1)&=\epsilon(L_2)=\frac{\epsilon(\sigma_{12})}{\sigma_{11}-\sigma_{22}}, \\
\epsilon(B_1)&=\epsilon(\phi)=\frac{\epsilon(\sigma_{23})}{\sigma_{22}-\sigma_{33}}, \\
\epsilon(B_2)&=\epsilon(\psi)=\frac{\epsilon(\sigma_{13})}{\sigma_{11}-\sigma_{33}}, \\
\epsilon^2(L_3)&=\frac{\phi^2\epsilon^2(\psi)+\psi^2\epsilon^2(\phi)}{(\phi^2+\psi^2)^2}, \\
\epsilon^2(B_3)&=\frac{\sin^2 L_3 \epsilon^2(\phi) + \cos^2 L_3 \epsilon^2(L_3)}{(\sin^2 L_3 + \phi^2)^2}.
\end{align}
\label{eq:LBsErrs}
\end{subequations}
Here
\begin{equation}
\phi = {\rm ctg}(B_3)\cos(L_3), \:\:\:
\psi = {\rm ctg}(B_3)\sin(L_3),
\label{eq:LBsErrs_2}
\end{equation}
and
\begin{subequations}
\begin{align}
\epsilon^2(\sigma_{12}) &= \frac{1}{N}\left(\frac{1}{N}\sum\limits_{k=1}^N V'^2_{x,k} V'^2_{y,k} - \sigma^2_{12}\right), \\
\epsilon^2(\sigma_{23}) &= \frac{1}{N}\left(\frac{1}{N}\sum\limits_{k=1}^N V'^2_{y,k} V'^2_{z,k} - \sigma^2_{23}\right), \\
\epsilon^2(\sigma_{13}) &= \frac{1}{N}\left(\frac{1}{N}\sum\limits_{k=1}^N V'^2_{x,k} V'^2_{z,k} - \sigma^2_{13}\right).
\end{align}
\label{eq:LBsErrs_3}
\end{subequations}

The uncertainties for $\sigma_\xi,\sigma_\eta, \sigma_\zeta$  were estimated using formulas 33 and 36 from \citet{Rootsmae1959}
\begin{subequations}
\begin{align}
\epsilon^2(\sigma_{\xi}) &= \frac{1}{4\sigma^2_{\xi}N}\left(\frac{1}{N}\sum\limits_{k=1}^N V'^4_{x,k} - \sigma^4_{x}\right), \\
\epsilon^2(\sigma_{\eta}) &= \frac{1}{4\sigma^2_{\eta}N}\left(\frac{1}{N}\sum\limits_{k=1}^N V'^4_{y,k} - \sigma^4_{y}\right), \\
\epsilon^2(\sigma_{\zeta}) &= \frac{1}{4\sigma^2_{\zeta}N}\left(\frac{1}{N}\sum\limits_{k=1}^N V'^4_{z,k} - \sigma^4_{z}\right).
\end{align}
\label{eq:diagSigmas_err}
\end{subequations}

This method allows us to estimate the uncertainties in determining the coordinates $L$, $B$ of the point of intersection of each of the axes $\xi$, $\eta$ and $\zeta$ of the ellipsoid with the celestial sphere in an independent way. The exceptions are the directions $L_1$ and $L_2$, the uncertainty of which coincide, since they are calculated using the same formula.

\citet{Amendt1991, Kuijken1991, Binney2008, Smith2012} showed that in a stationary axisymmetric disk galaxy (Oort rotation) the ellipsoid axes ideally coincide with the Galactic coordinate axes. In addition, \citet{Dehnen2000, Vorobyov2008, Minchev2010, Saha2013} showed that non-axisymmetric structures can have a noticeable effect on the observed orientation of stellar velocity ellipsoids. Therefore, the discrepancy between the $\xi$ axis of an ellipsoid directed to the vertex point $L$, $B$ and the Galactic $X$ axis is usually interpreted as a deviation from the Oort rotation. When determining the orientation of the velocity ellipsoid relative to the local Galactic coordinate system, the deviation of the local longitude and local latitude of the vertex point from the direction to the Galactic center ($L_v-L_{GC}$) and ($B_v-B_{GC}$) is considered. 
In this case, the vertex is understood as a point in the sky relative to which the centroids move at circular orbits. The Galactic center is understood as a point on the celestial sphere with coordinates $\alpha_{GC}$ = 266$^\circ$.40499, $\delta_{GC} = -28^\circ$.93617 adopted by the Hipparcos consortium \citet{Perryman1997}, that corresponds to $L_{GC} = 0^\circ$ and $B_{GC} = 0^\circ$. In general, the velocity ellipsoid has an arbitrary shape, reflecting the anisotropy in the motion of stars and the arbitrary orientation of the directions of the greatest and least velocity dispersion, as a result of which the axes of the ellipsoid do not coincide with the Galactic coordinate axes.

\section{Results}
\label{sec:results}
\subsection{Determination of the coordinates of the intersection points of the velocity ellipsoid axes with the celestial sphere.}

We analyze orientations of the velocity ellipsoids determined from the spatial velocities of the stars. To do this, we find the coordinates of the points on the celestial sphere $L$ and $B$ in the local coordinate system, towards which the ellipsoids' axes are directed. It is obvious that the angle between the local $x$-axis and the largest axis of velocity ellipsoid $\xi$, directed to the vertex, is the difference between the local longitudes of the Galactic center and the vertex. 
Since the Galactic center coordinates in any local Galactic coordinate system are equal to zero, the longitude of the vertex in this system will always be numerically equal to the deviation of the vertex in longitude $Lv=L_\xi$, and the latitude to the deviation of the vertex in latitude $Bv=B_\xi$.
We determine the direction of the $x$-axis of the local system from an arbitrary centroid to the Galactic center, as well as its Galactocentric coordinates $R$ and $\theta$, assuming that the Galactocentric distance of the Sun $R_\odot$ is 8.28 kpc \citep[][]{Abuter2021}.

\begin{figure*}
    \centering
    \includegraphics[width=0.45\linewidth]{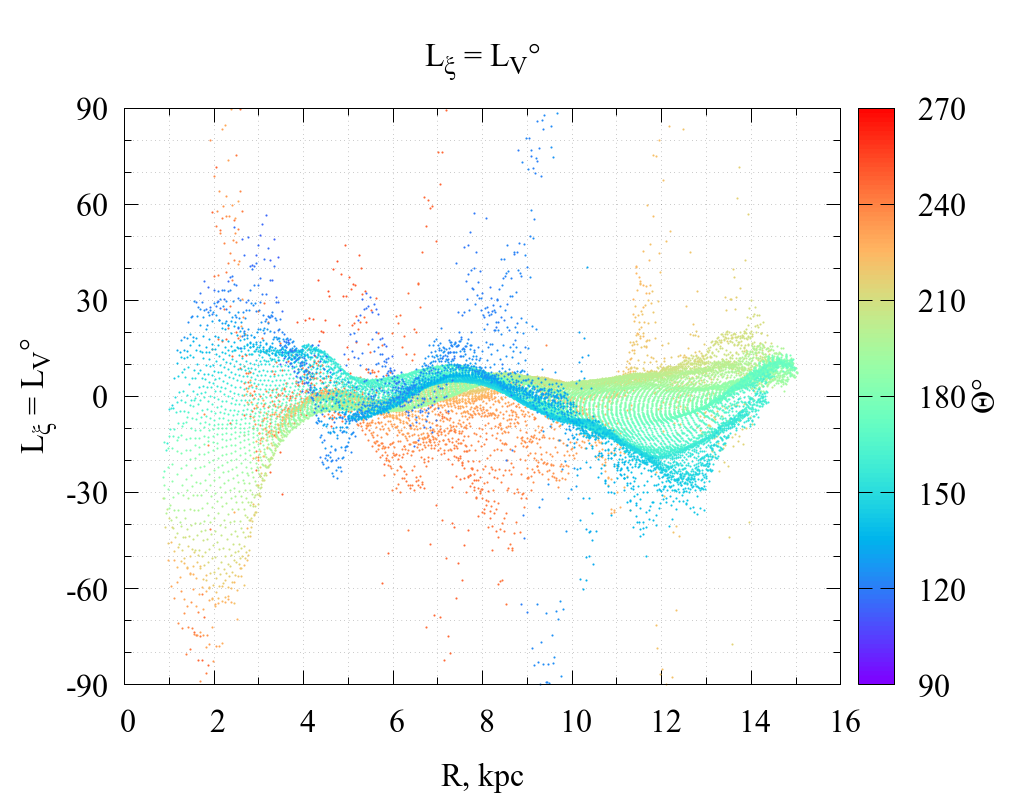}
    \includegraphics[width=0.45\linewidth]{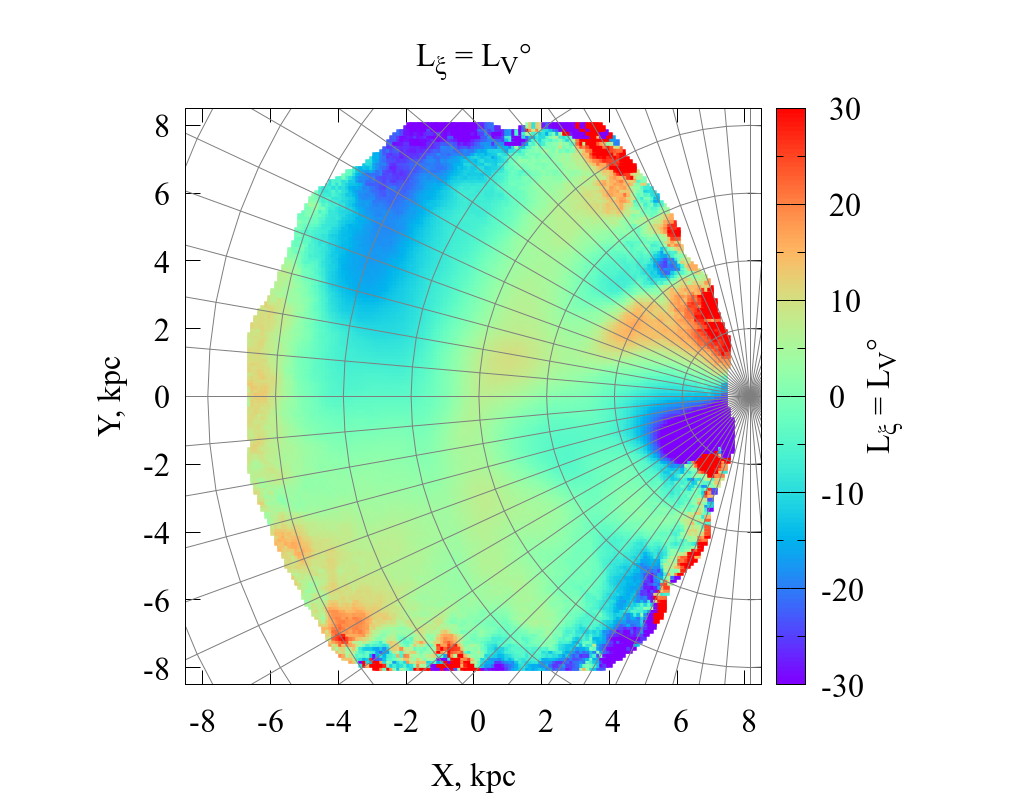}
    \caption{Left: the angle $L_\xi$ depending on the Galactocentric distance $R$, with $\theta$ shown in color. Right: the distribution map of $L_\xi$ in the rectangular Galactocentric coordinates $XY$.}
    \label{fig:L1}
\end{figure*}

\begin{figure*}
    \centering
    \includegraphics[width=0.45\linewidth]{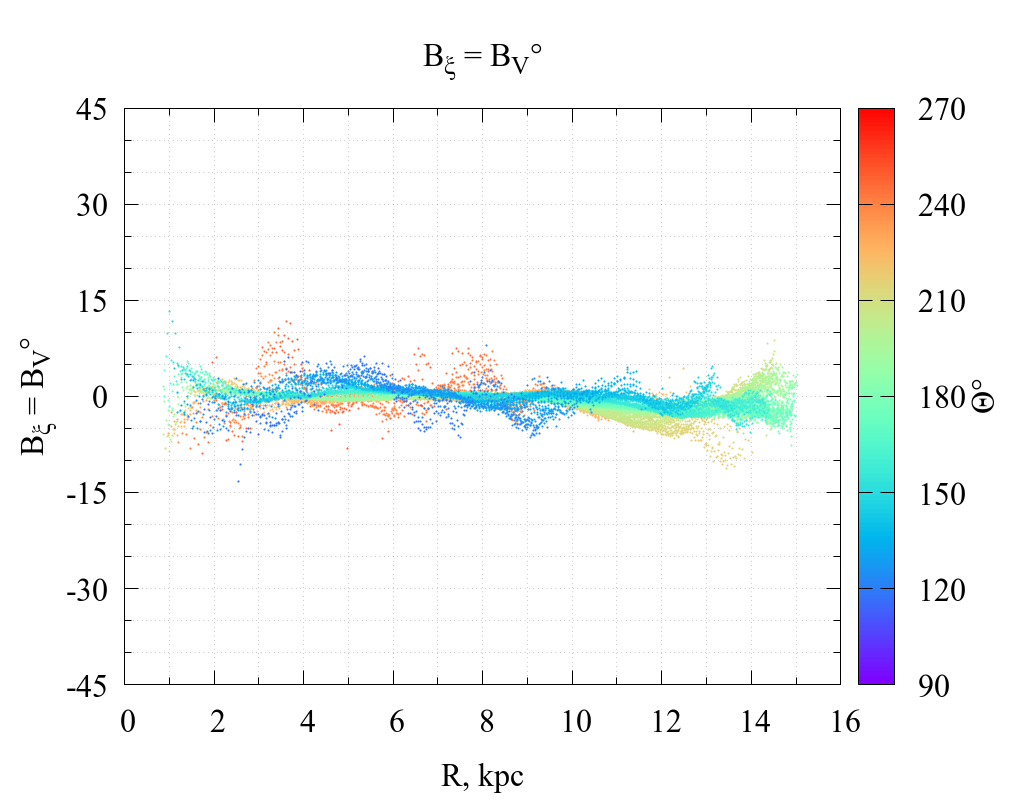}
    \includegraphics[width=0.45\linewidth]{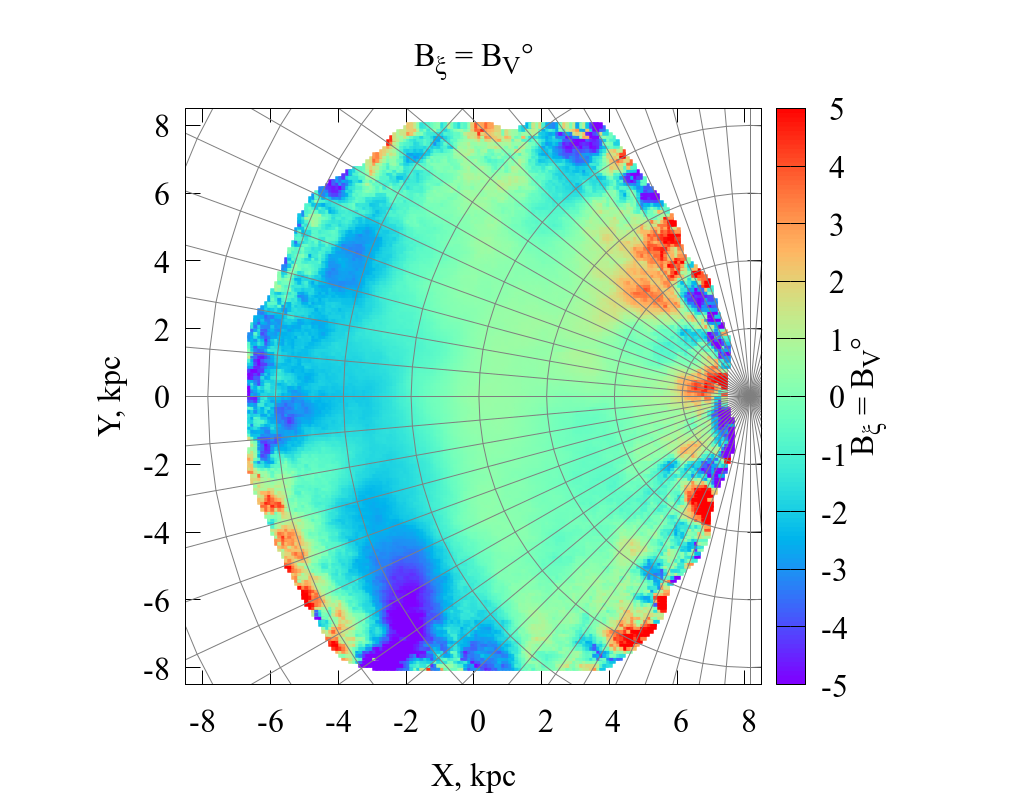}
    \caption{Left: the angle $B_\xi$ depending on the Galactocentric distance $R$, with $\theta$ shown in color. Right: the distribution map of $B_\xi$ in the rectangular Galactocentric coordinates $XY$.}
    \label{fig:B1}
\end{figure*}

\begin{figure*}
    \centering
    \includegraphics[width=0.45\linewidth]{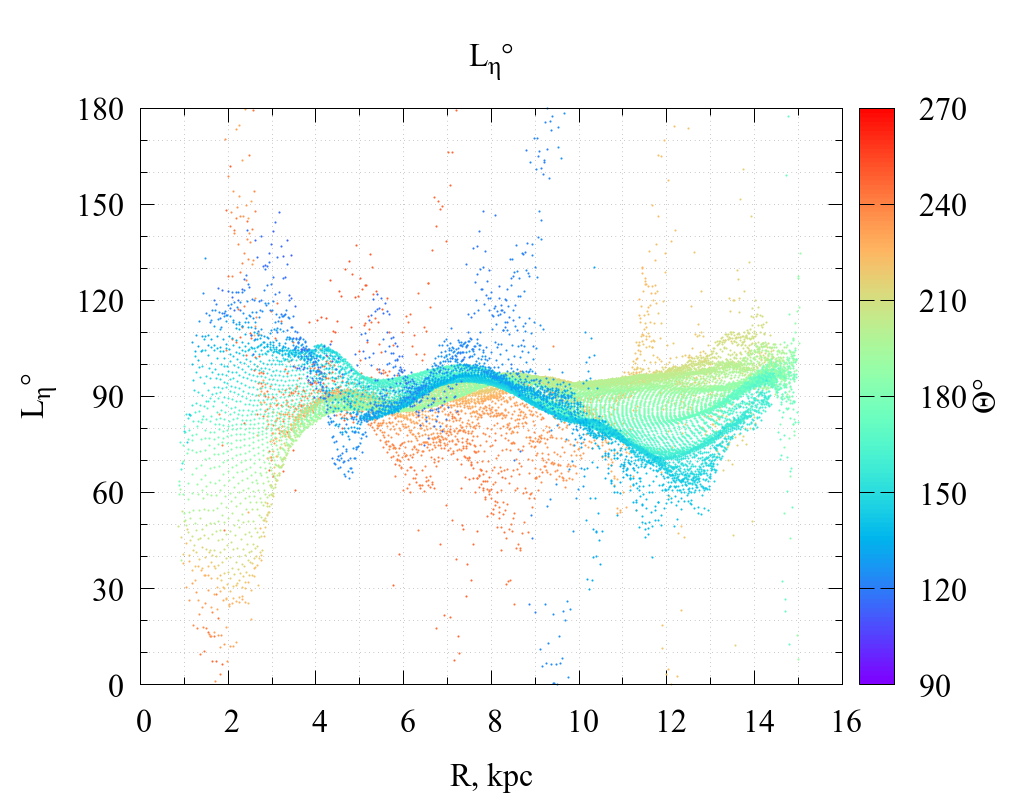}
    \includegraphics[width=0.45\linewidth]{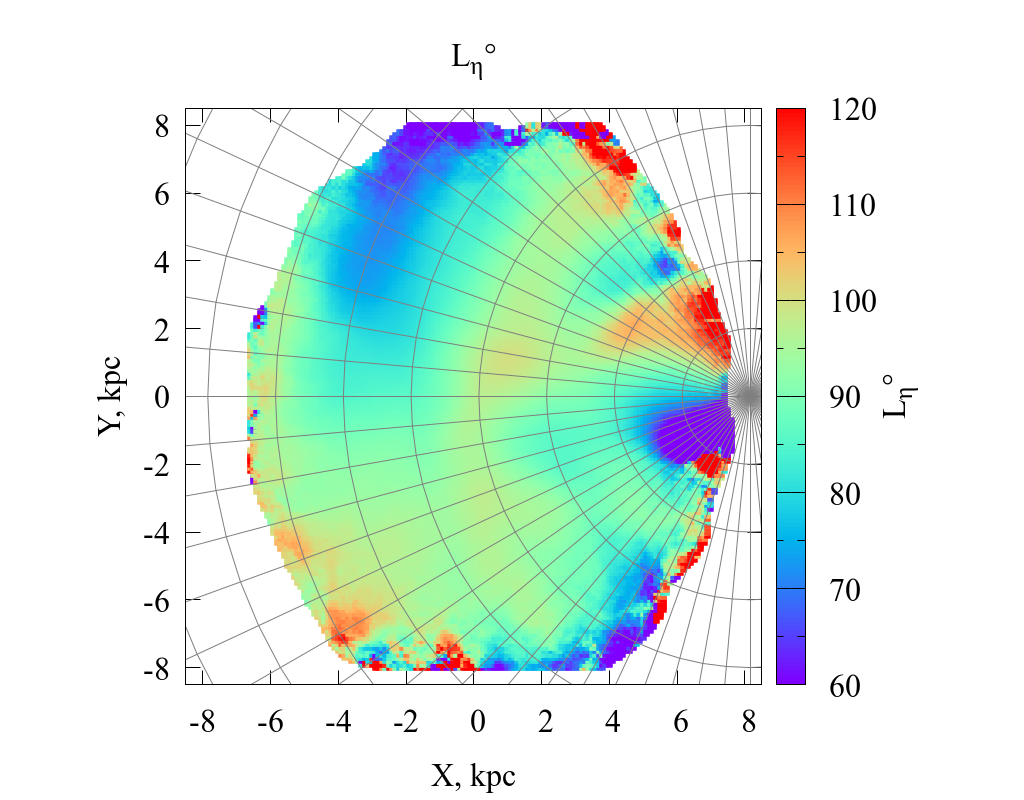}
    \caption{Left: the angle $L_\eta$ depending on the Galactocentric distance $R$, with $\theta$ shown in color. Right: the distribution map of $L_\eta$ in the rectangular Galactocentric coordinates $XY$.}
    \label{fig:L2}
\end{figure*}

\begin{figure*}
    \centering
    \includegraphics[width=0.45\linewidth]{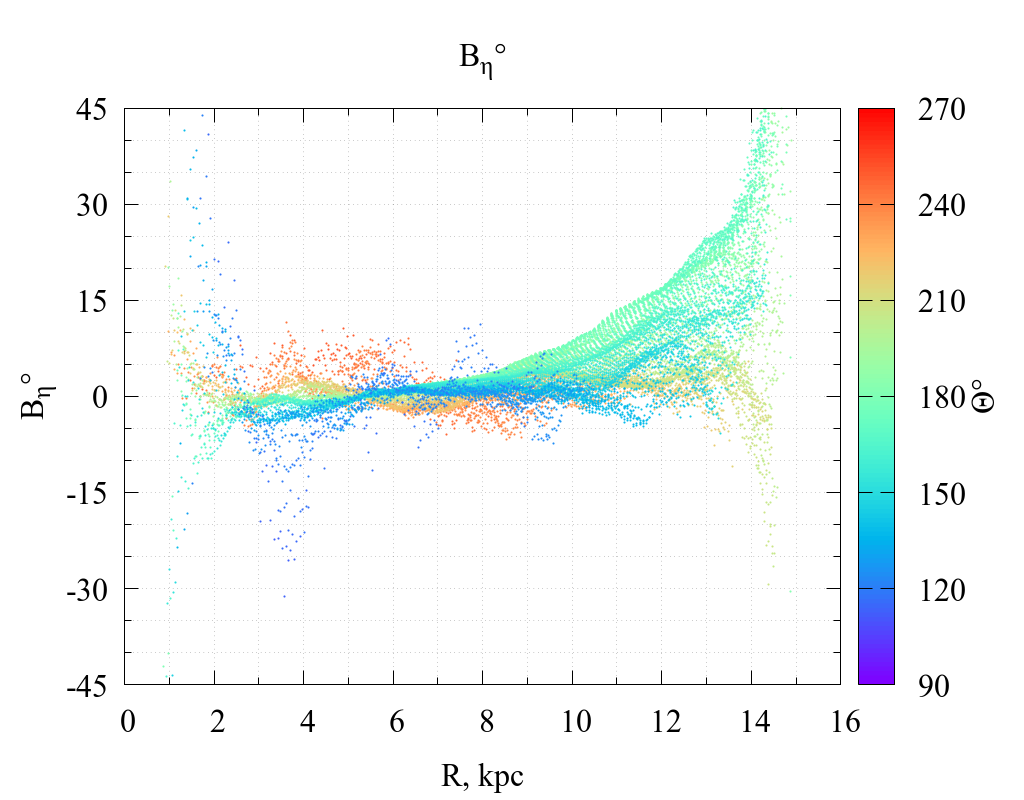}
    \includegraphics[width=0.45\linewidth]{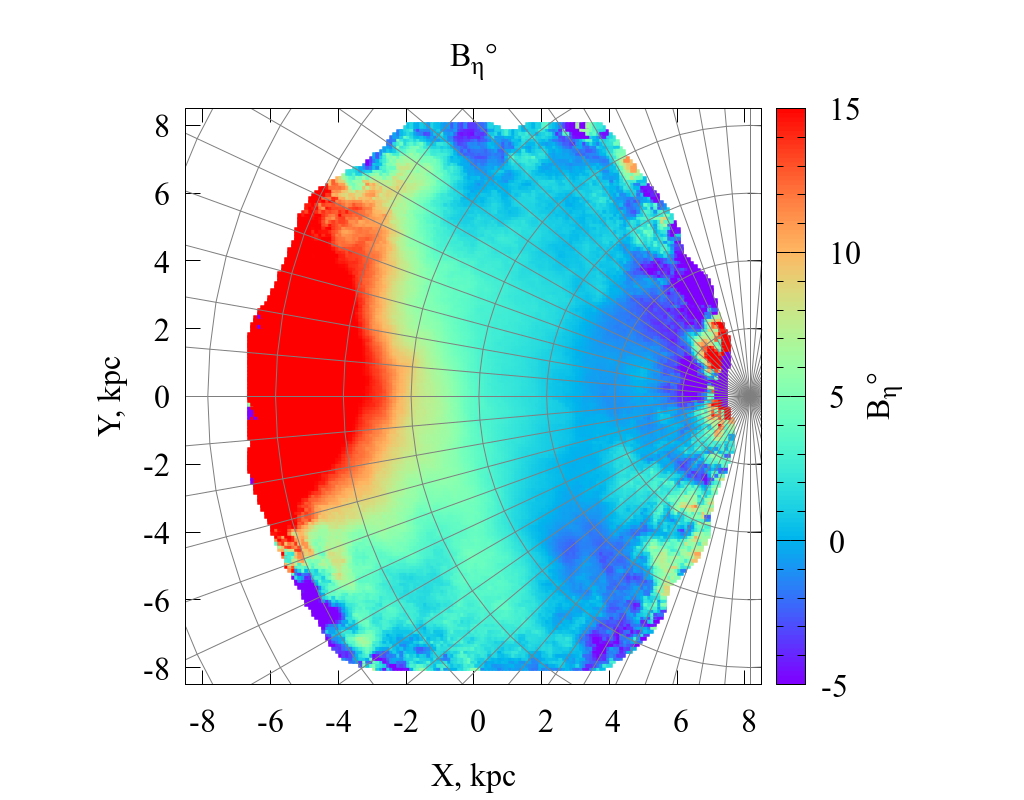}
    \caption{Left: the angle $B_\eta$ depending on the Galactocentric distance $R$, with $\theta$ shown in color. Right: the distribution map of $B_\eta$ in the rectangular Galactocentric coordinates $XY$.}
    \label{fig:B2}
\end{figure*}

\begin{figure*}
    \centering
    \includegraphics[width=0.45\linewidth]{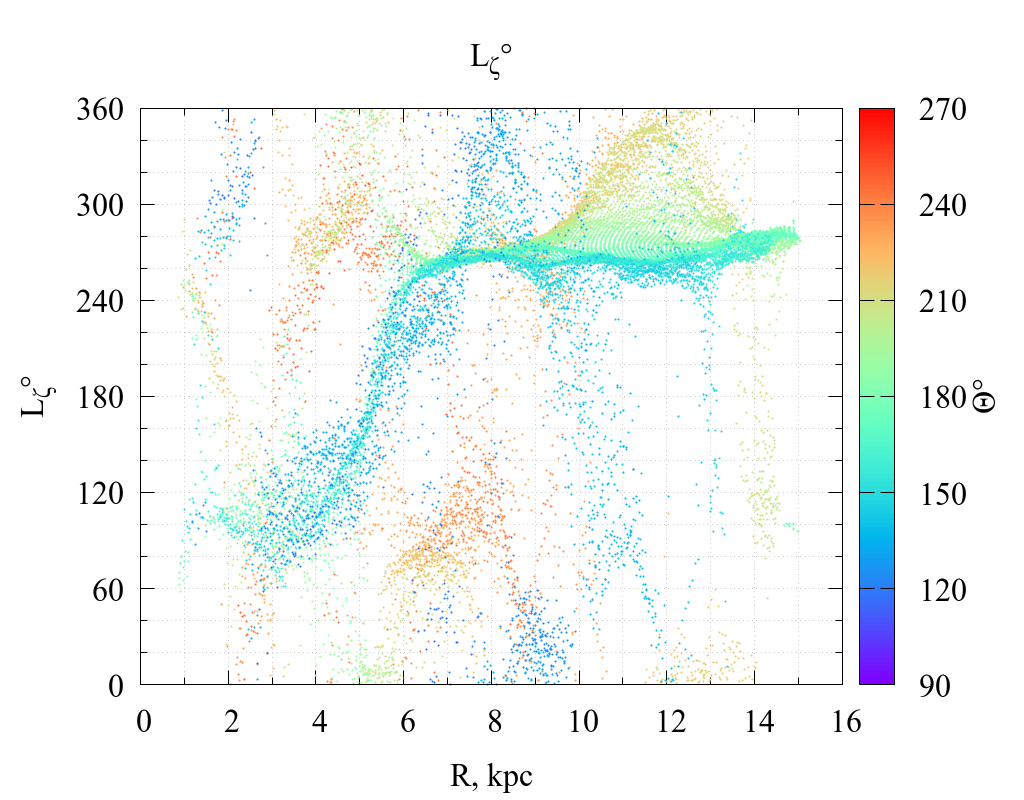}
    \includegraphics[width=0.45\linewidth]{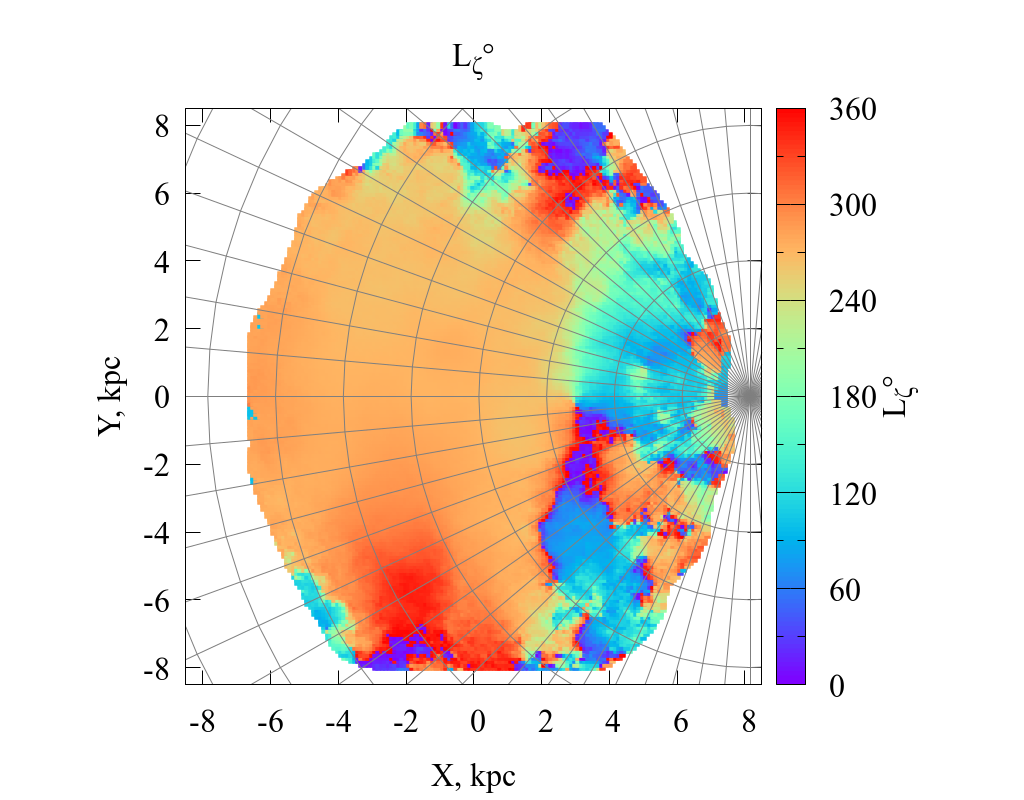}
    \caption{Left: the angle $L_\zeta$ depending on the Galactocentric distance $R$, with $\theta$ shown in color. Right: the distribution map of $L_\zeta$ in the rectangular Galactocentric coordinates $XY$.}
    \label{fig:L3}
\end{figure*}

\begin{figure*}
    \centering
    \includegraphics[width=0.45\linewidth]{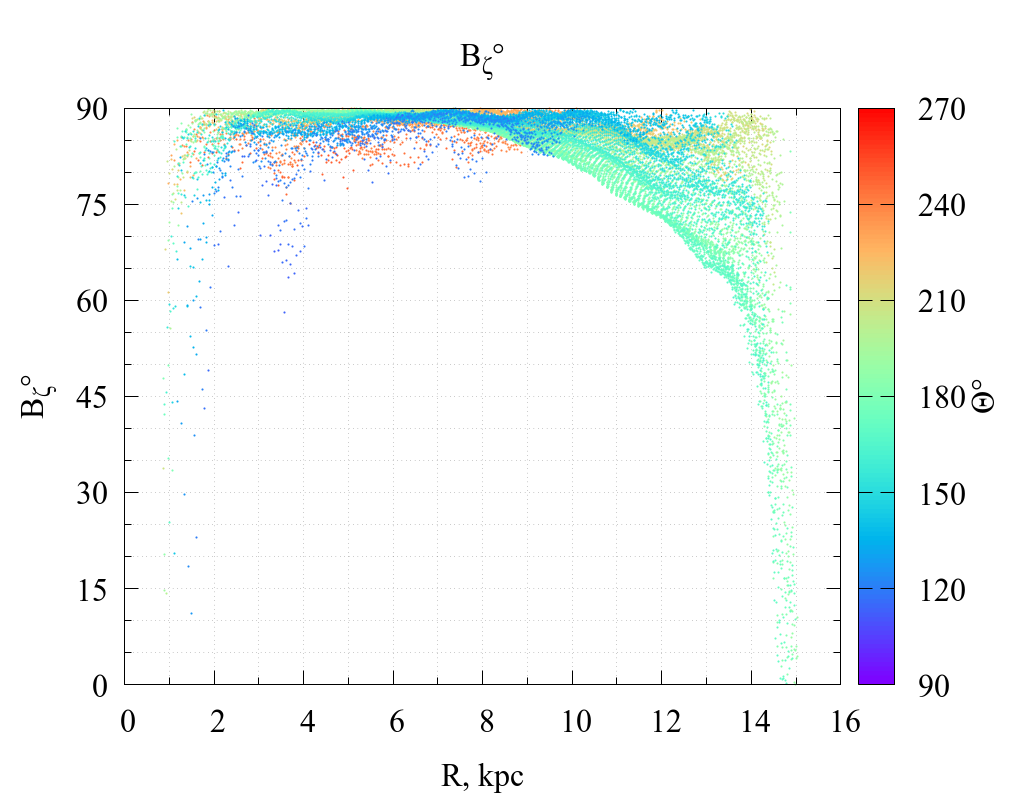}
    \includegraphics[width=0.45\linewidth]{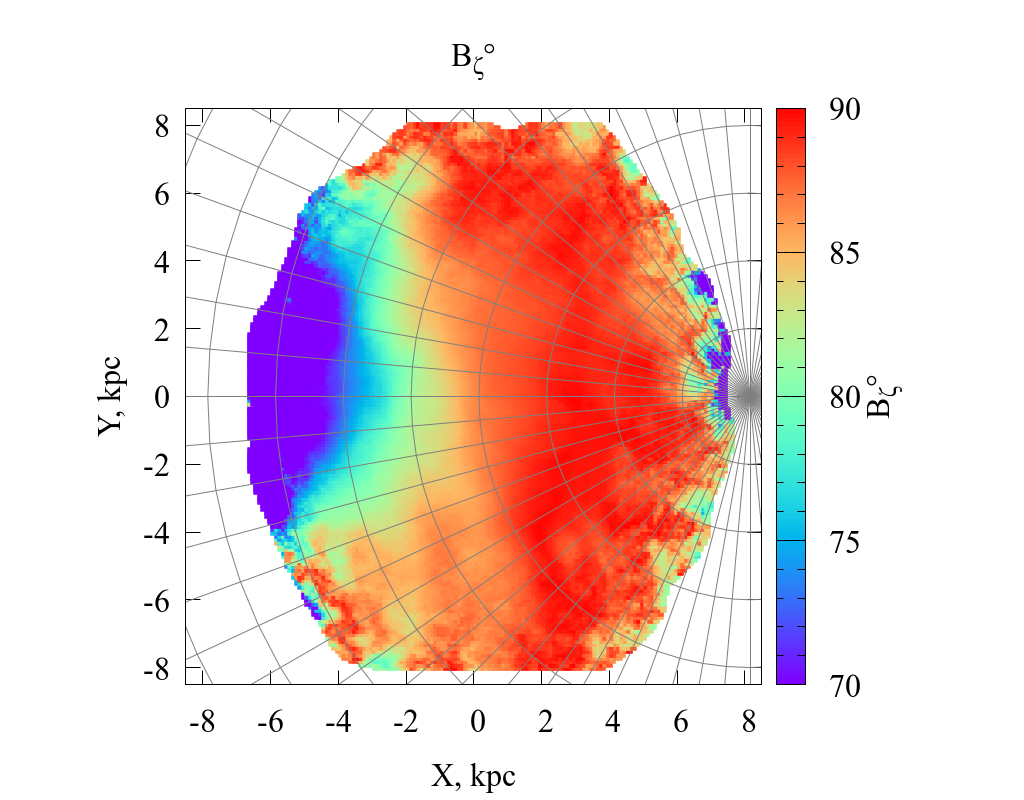}
    \caption{Left: the angle $B_\zeta$ depending on the Galactocentric distance $R$, with $\theta$ shown in color. Right: the distribution map of $B_\zeta$ in the rectangular Galactocentric coordinates $XY$.}
    \label{fig:B3}
\end{figure*}

The figures \ref{fig:L1}-\ref{fig:B3} show the coordinates $L$ and $B$ of the first, second and third axes of the velocity ellipsoids (in this order) depending on the Galactocentric distance $R$ and the azimuthal angle $\theta$, as well as their distribution in the Galactic mid-plane. The uncertainties in determining the coordinates are given in Appendix \ref{sec:sigma_LB} (Figs. \ref{fig:eL1}-\ref{fig:eB3}).
In Fig. \ref{fig:L1} on the left, the “stratification and interweaving” of the dependencies $L_\xi (R)$ corresponding to different angles $\theta$ are clearly visible. Thus, the angles $L_\xi$ corresponding to the velocity ellipsoids located in the range $150^\circ < \theta < 170^\circ $ have a value and the nature of their dependence on $R$ that are noticeably different from those in the range $ 200^\circ < \theta < 220^\circ$. This indicates that the stellar regions under study do not follow an axisymmetric rotation.
The right panel of Fig. \ref{fig:L1} shows that the angle $L_\xi$ is not equal to zero, but noticeably changes in the Galactic plane $XY$. The concentric zones between the radii of 6 and 8 kpc in the region of $90^\circ<\theta<180^\circ$ and between the radii of 8 and 10 kpc in the range $180^\circ<\theta<270^\circ$ are especially prominent. Here, the deviations of the vertices reach $15^\circ-20^\circ$. The direction $\theta\sim180^\circ$ is a conditional symmetry line, with respect to which the sign of $L_\xi$ is mainly different. This is especially seen at large deviations from the angle $\theta=180^\circ$.

Left panel of Fig. \ref{fig:B1} shows that in most cases (up to distances of about 10 kpc) the angle $B_\xi$ weakly depends on the Galactocentric coordinates.
Only at distances of about 10 kpc it systematically increases and at a distance of about 13 kpc it reaches a value of about $\sim3^\circ-4^\circ$. The angle $B_\xi$ demonstrates the same behavior in the right panel of Fig. \ref{fig:B1}. Here, it is clearly seen that the first largest axes of the ellipsoids $\xi$ practically do not deviate from the Galactic mid-plane up to distances of about 10 kpc. And at greater distances they incline by an angle of about $\sim3^\circ-4^\circ$.
In Figs. \ref{fig:eL1}, \ref{fig:eB1} in Appendix \ref{sec:sigma_LB}, the uncertainties in determining $L_\xi$ and $B_\xi$ generally do not exceed 1.0$^\circ$ up to $R$ of 12 kpc, and at higher distances they begin to increase and reach a value of $\sim4.0^\circ$.

Fig. \ref{fig:L2}shows the distribution in the galactic plane of the longitudes $L_\eta$ of the second axis of the ellipsoids $\eta$. It is clearly seen that the difference in the longitudes of the axes $\eta$ and $y$ exactly repeats the behavior of the difference in the longitudes of the axes $\xi$ and $x$, which naturally leads to the distribution identity of the deviations of the longitudes relative to the directions $x$ and $y$.

The behavior of the angles $B_\eta$ looks different from $B_\xi$(see Fig. \ref{fig:B2}).
As can be seen from the figure, for ellipsoids located at distances between $\sim5$ kpc and $\sim8$ kpc, the angle $B_\eta$ changes little ($\sim3^\circ-4^\circ$), and then increases sharply reaching $\sim20^\circ$ at a distance of 13 kpc. 
In the range of azimuthal angles of $150^\circ<\theta<210^\circ$, starting from distances of about 8 kpc, the $\eta$ axis deviates more and more from the Galactic mid-plane and at a distance of $\sim14$ kpc exceeds $\sim20^\circ$. This behavior of the latitude values for the first and second axes of the ellipsoids indicates that the planes of the ellipsoids determined by the $\xi$ and $\eta$ axes begin to deviate more and more from the Galactic mid-plane with increasing distance $R$, while maintaining the ratio $B_\eta/B_\xi>1$. 
These observed tilts of the $\xi$ and $\eta$ axes clearly indicate a kinematic deformation of the velocity field that begins at approximately the Solar circle. If we assume that the observed deformations of the velocity field are associated with the Galactic warp, the angles $B_\xi$ and $B_\eta$ can be considered as its kinematic signatures.

The uncertainties in determining the angles $L_\eta$ and $B_\eta$ (see Figs. \ref{fig:eL2}, \ref{fig:eB2} in Appendix \ref{sec:sigma_LB}) are practically the same as for the angles $L_\xi$ and $B_\xi$. At distances over 13 kpc they increase sharply and reach tens of degrees. This is apparently due to the accuracy of determining the parallaxes.

Fig. \ref{fig:L3} and \ref{fig:B3} show the coordinates $L_\zeta$ and $B_\zeta$ of the third axis of the ellipsoids $\zeta$ depending on the Galactocentric distance $R$ and the azimuthal angle $\theta$, as well as their distribution in the Galactic plane. The scatter of the coordinate $L_\zeta$ is large, which is quite expected, because the uncertainty of the longitude $L_\zeta$ value increases when approaching the latitude value $B_\zeta=90^\circ$ (see Figs. \ref{fig:L3} and \ref{fig:B3}). It is also clearly seen that starting from distances greater than 6 kpc, the values of the longitudes $L_\zeta$ are grouped near 270$^\circ$ ($-90^\circ$), which is consistent with the values of the deviations in the latitude $B_\eta$. 

We did not find an unambiguous explanation for the behavior of $L_\zeta$ at distances less than 6 kpc. However, it can be assumed that the Galactocentric distance of 6 kpc is some special point. 
It is clearly seen in Fig. \ref{fig:L1} that in the range from 3 to 6 kpc there is a rotation of the ellipsoid plane $\xi\eta$ around the galactic axis $X$. In other words, there is a change in the sign of the tilt of the plane $\xi\eta$ relative to the Galactic plane. The change in the sign of the tilt is also confirmed by the behavior of the angle $B_\eta$ depending on the Galactocentric coordinates $R$ $\theta$. And starting from the point $R>6$kpc, the rotation of the ellipsoid plane $\xi\eta$ around the galactic axis $X$ disappears, and the longitudes $L_\zeta$ remain almost unchanged. Also, at a distance of about 6 kpc we see the beginning of a change in the behavior of the dispersions $\sigma_\xi, \sigma_\eta, \sigma_\zeta$ (see Figs. \ref{fig:Sq_1} - \ref{fig:Sq_3}) depending on $R$.

It is obvious that the value of $B_\zeta$ will always be in the range from $0^\circ$ to $+90^\circ$. As can be seen in Fig. \ref{fig:B3}, the value of the latitude of the third axis (angle $B_\zeta$) in the range of azimuthal angles $150^\circ<\theta<210^\circ$ and distances less than 10 kpc is practically equal to $+90^\circ$. Outside these ranges, the increase in deviation from $+90^\circ$ is quite obvious, reaching ten or more degrees at a distance of more than 12 kpc, which is in excellent agreement with the behavior of the axes $\xi$ and $\eta$ and confirms the correctness of the results obtained.
The uncertainties in determining the angle $B_\zeta$ (see Fig. \ref{fig:eB3} in Appendix \ref{sec:sigma_LB}) are practically the same as for $B_\xi$ and $B_\eta$. And for the angle $L_\zeta$ (Figs. \ref{fig:eL3}, ), their smallest value is approximately 2 degrees in the range of distances between 7 kpc and 13 kpc, which is noticeably narrower than for other axes.

\subsection{Determining the shape of velocity ellipsoids}

The shape of the velocity ellipsoid reflects the anisotropy in the stellar motions. It can be used as a kinematic characteristic of stars and characterized by the ratio of the lengths of the ellipsoid semi-axes or the velocity dispersions along the directions of the corresponding axes.

\begin{figure*}
    \centering
    \includegraphics[width=0.45\linewidth]{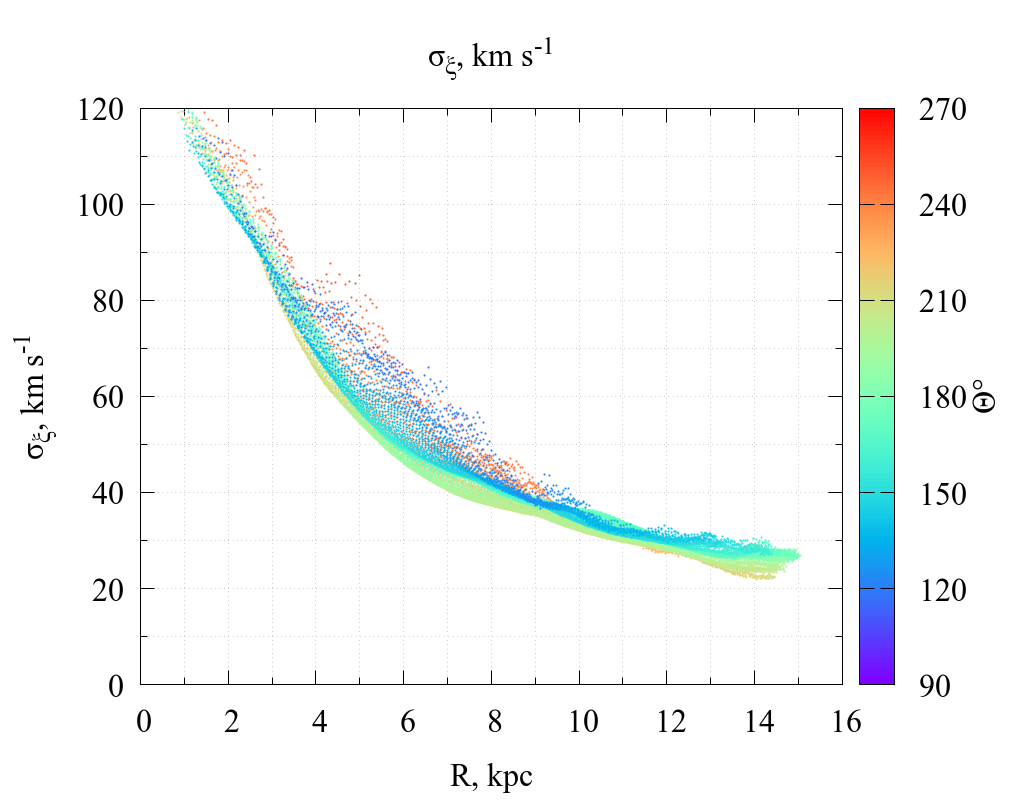}
    \includegraphics[width=0.45\linewidth]{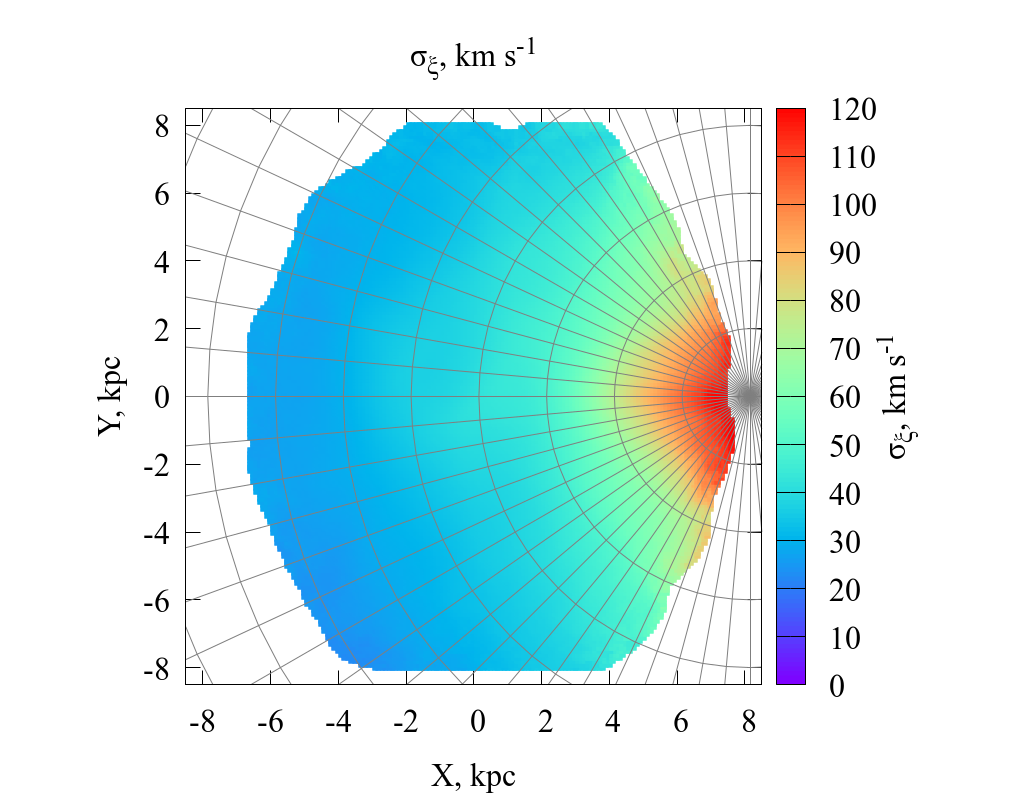}
    \caption{Left: the velocity ellipsoid semi-axis lengths $\sigma_\xi$ depending on the Galactocentric distance $R$, with $\theta$ shown in color. Right: the distribution map of $\sigma_\xi$ on the rectangular Galactocentric coordinates $XY$.}
    \label{fig:Sq_1}
\end{figure*}

\begin{figure*}
    \centering
    \includegraphics[width=0.45\linewidth]{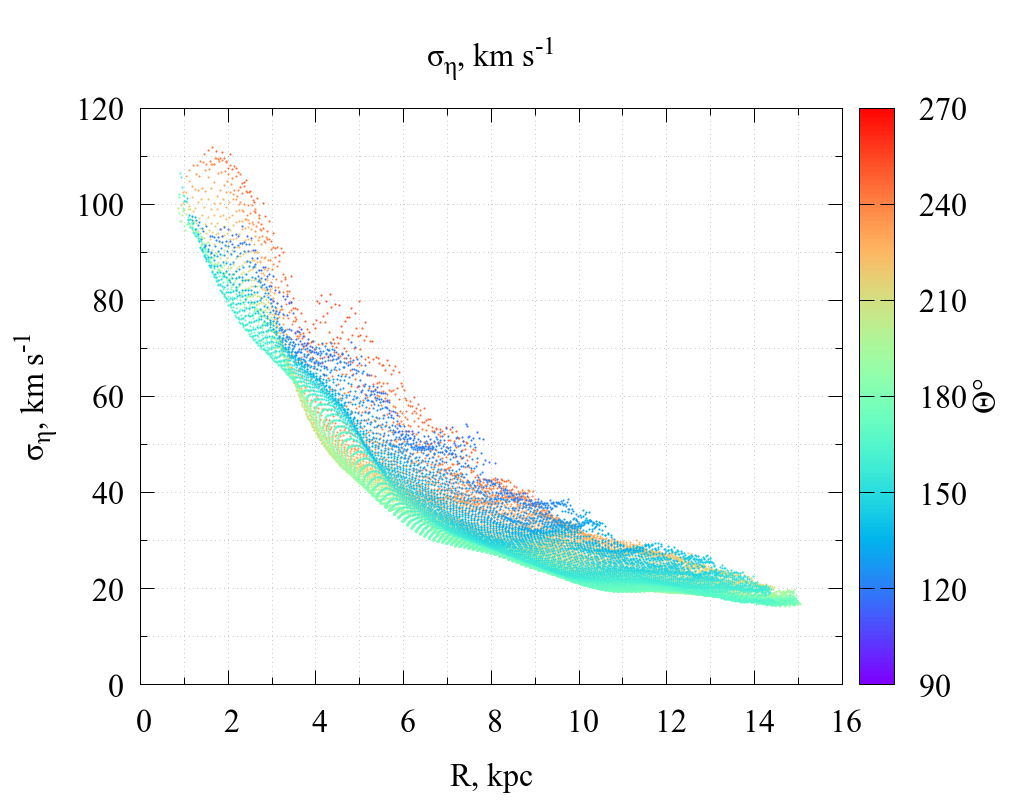}
    \includegraphics[width=0.45\linewidth]{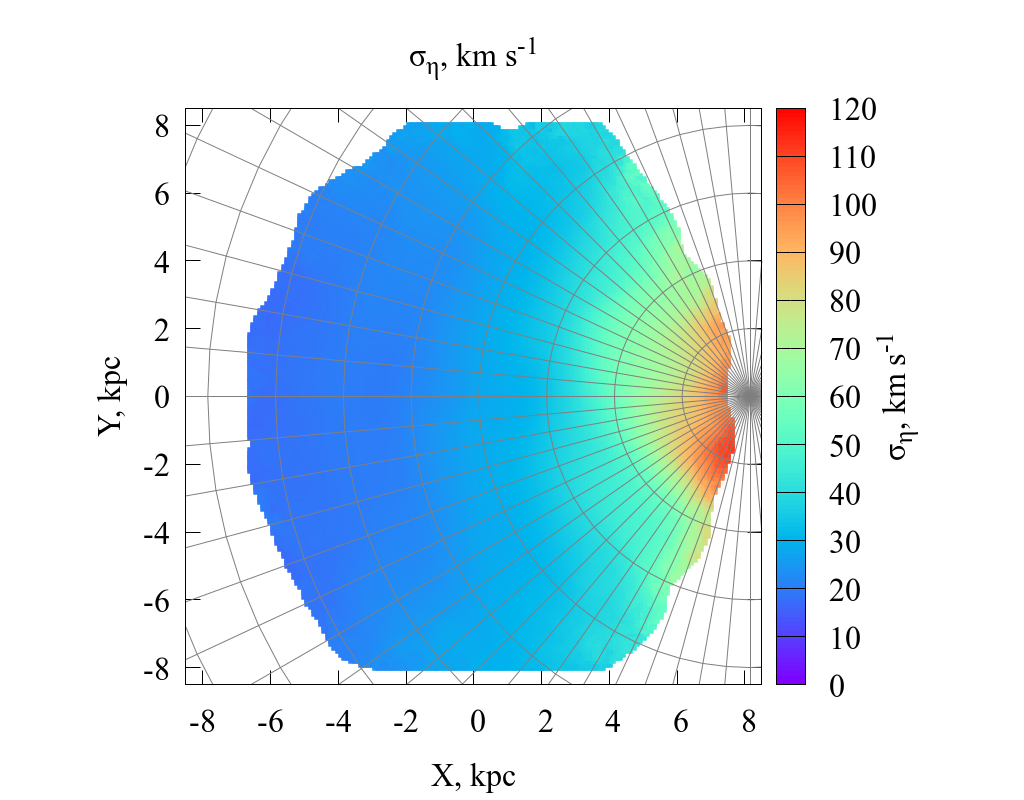}
    \caption{Left: the velocity ellipsoid semi-axis lengths $\sigma_\eta$ depending on the Galactocentric distance $R$, with $\theta$ shown in color. Right: the distribution map of $\sigma_\eta$ in the rectangular Galactocentric coordinates $XY$.}
    \label{fig:Sq_2}
\end{figure*}

\begin{figure*}
    \centering
    \includegraphics[width=0.45\linewidth]{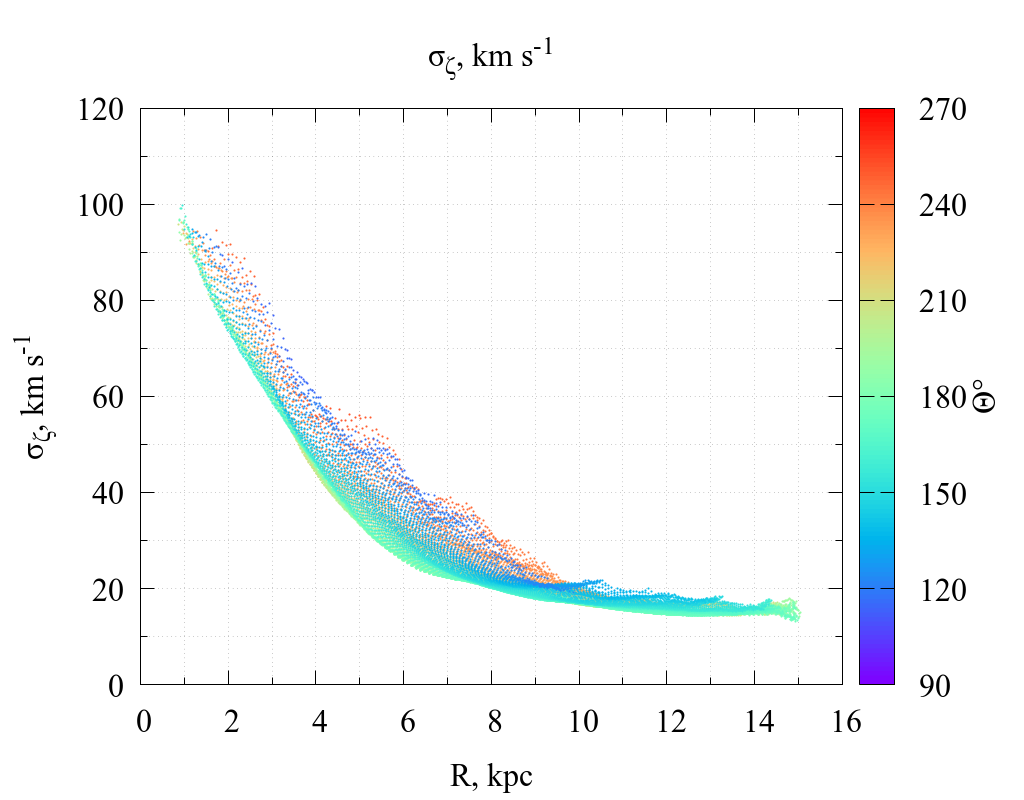}
    \includegraphics[width=0.45\linewidth]{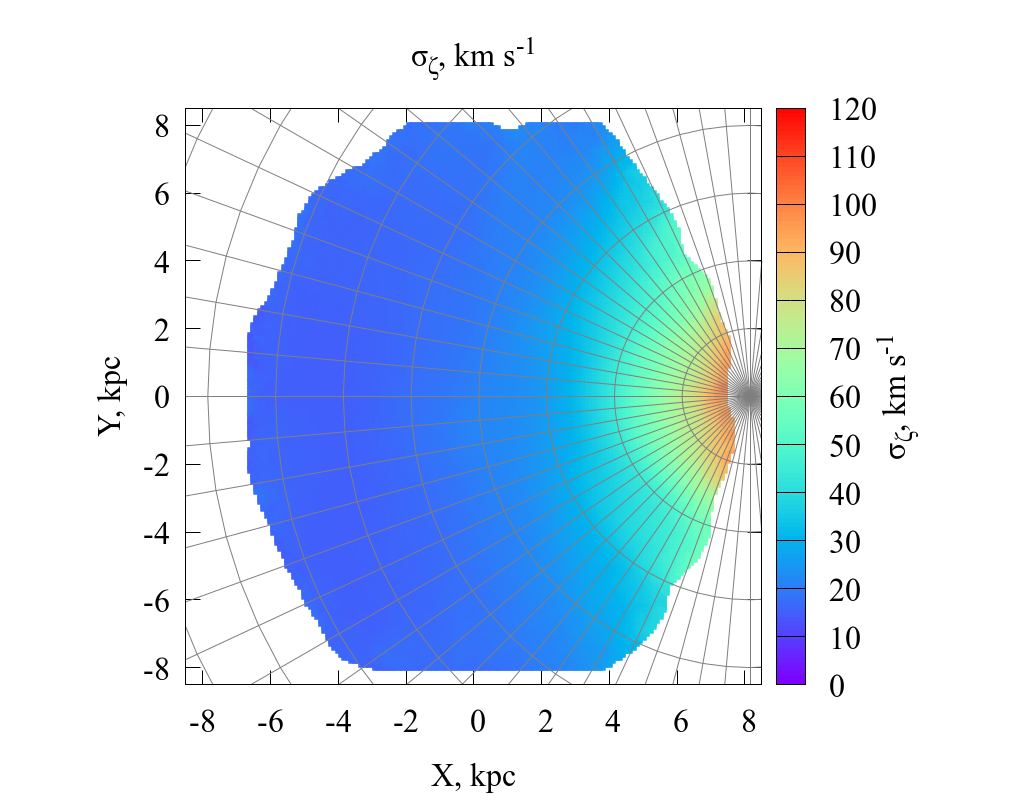}
    \caption{Left: the velocity ellipsoid semi-axis lengths $\sigma_\zeta$ depending on the Galactocentric distance $R$, with $\theta$ shown in color. Right: the distribution map of $\sigma_\zeta$ in the rectangular Galactocentric coordinates $XY$.}
    \label{fig:Sq_3}
\end{figure*}

Figs. \ref{fig:Sq_1}-\ref{fig:Sq_3} show the dependencies of the ellipsoid semi-axis lengths on the Galactocentric coordinates $R$ and $\theta$ (left panel) and their distribution $\sigma_\xi, \sigma_\eta, \sigma_\zeta$ in the Galactic plane $XY$ (right panel). As can be seen from the figures, the lengths of the semi-axes depend significantly on the Galactocentric distance $R$. The maximum slope of the dependencies is observed in the range $0-6$kpc, and then the slope decreases with $R$ so that at $R=14$ kpc the values of $\sigma_\xi, \sigma_\eta, \sigma_\zeta$ are approximately equal to 26.5, 17.2, 14.7 km s$^{-1}$, respectively, with uncertainties of several km s$^{-1}$. The level and behavior of $\sigma_\xi, \sigma_\eta, \sigma_\zeta$ are apparently related to the existence of gravitational disturbances from the central part of the Galaxy in the range of $R$ from $0$ to $6$kpc, the influence of which decreases with increasing Galactocentric distance. And already at distances of approximately 6 kpc and further we observe a more ordered stellar motions, manifested in decreasing velocity dispersions. The dependence on $\theta$ is observed in the form of stratification, with $\sigma_\eta$ having the greatest scatter. At the distance of the Sun it reaches a value of $\sim20$ km s$^{-1}$.

\begin{figure*}
    \centering
    \includegraphics[width=0.45\linewidth]{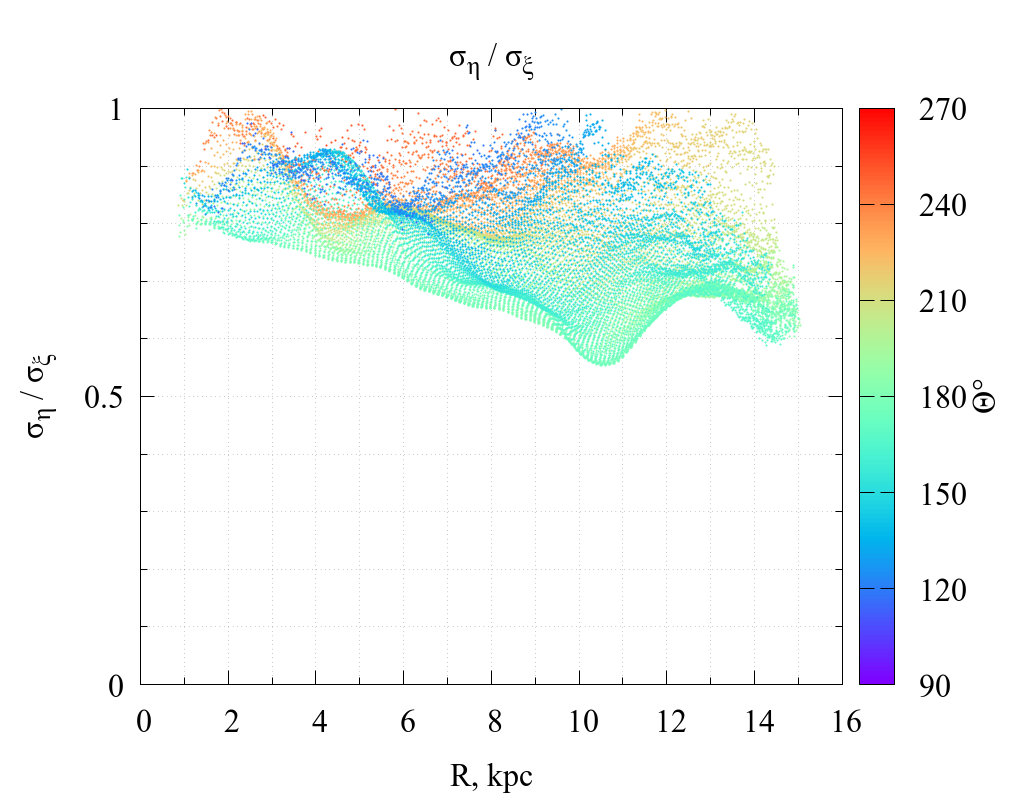}
    \includegraphics[width=0.45\linewidth]{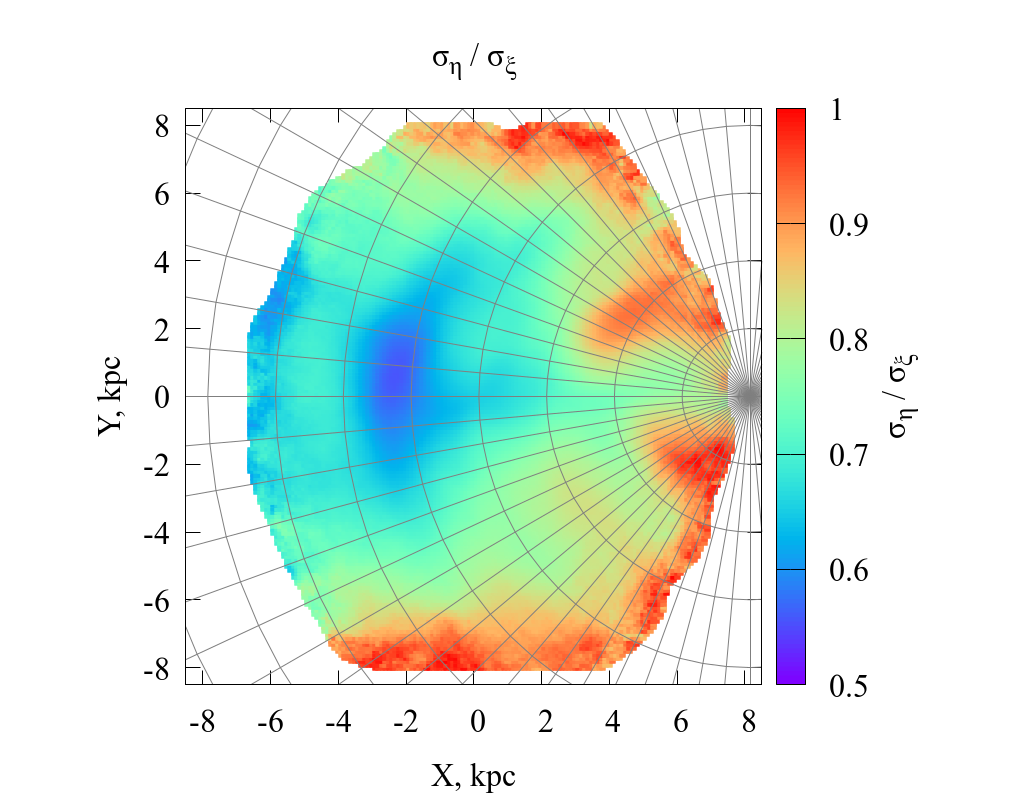}
    \caption{Left: the ratio $\sigma_\eta/\sigma_\xi$ depending on the Galactocentric distance $R$, with $\theta$ shown in color. Right: the distribution map of $\sigma_\eta/\sigma_\xi$ in the rectangular Galactocentric coordinates $XY$.}
    \label{fig:div_21}
\end{figure*}

\begin{figure*}
    \centering
    \includegraphics[width=0.45\linewidth]{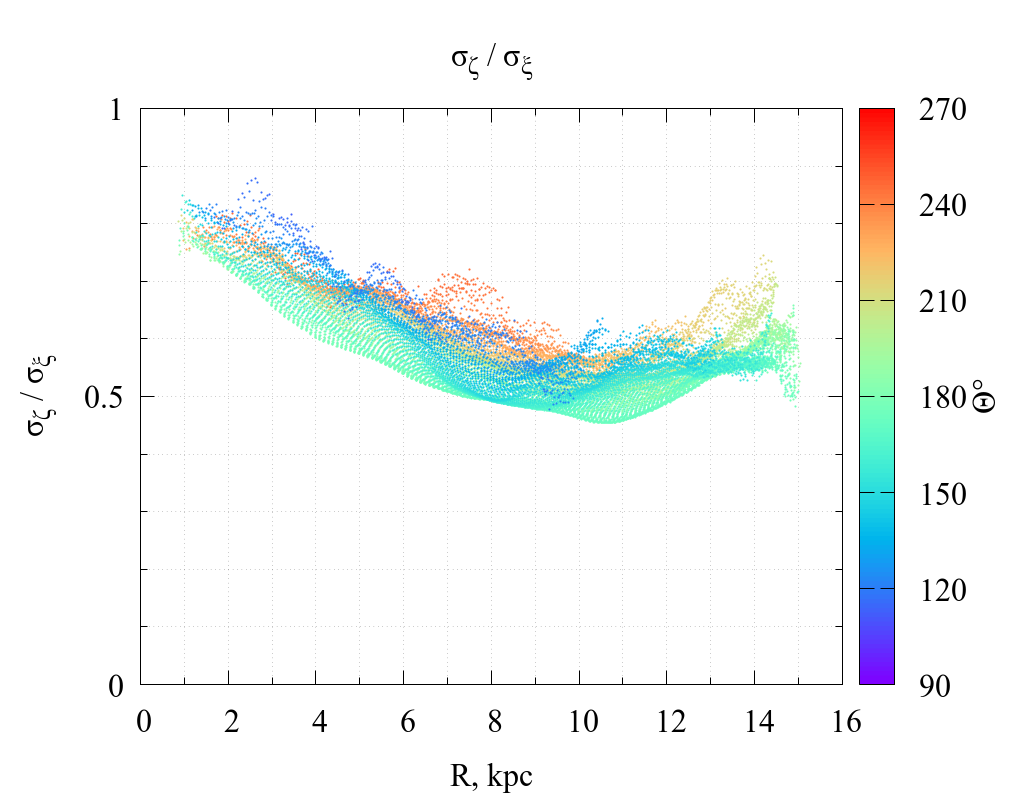}
    \includegraphics[width=0.45\linewidth]{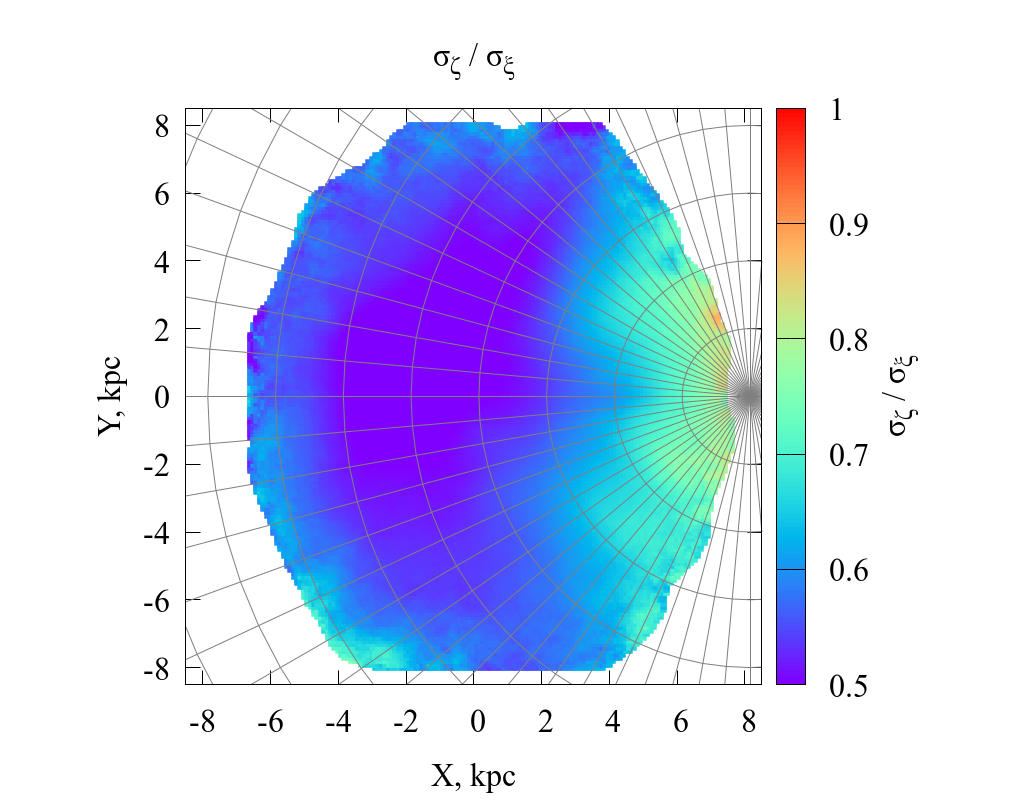}
    \caption{Left: the ratio $\sigma_\zeta/\sigma_\xi$ depending on the Galactocentric distance $R$, with $\theta$ shown in color. Right: the distribution map of $\sigma_\zeta/\sigma_\xi$ in the rectangular Galactocentric coordinates $XY$.}
    \label{fig:div_31}
\end{figure*}

Figs. \ref{fig:div_21}, \ref{fig:div_31} show the distributions of the semi-axis ratios. The ratio $\lambda_1>\lambda_2>\lambda_3$ is maintained. It is also clearly seen that the maximum difference in the lengths of the semi-axes $\sigma_\xi, \sigma_\eta$ is in the cone of azimuthal angles $165^\circ<\theta<185^\circ$, and the ratio $\sigma_\eta/\sigma_\xi$ approaches unity with increasing azimuth. In the region of $R\sim11$ kpc and $165^\circ<\theta<185^\circ$ there is a certain feature, which is also confidently observed in the ratio $ \sigma_\zeta/\sigma_\xi$. In this region, the lengths of the semi-axes the difference between $\sigma_\xi$ and $\sigma_\eta$ is about 1.5 times, and $\sigma_\xi$ and $\sigma_\zeta$ is almost 2 times. A distinct dependence of $\sigma_\eta/\sigma_\xi$ on $R$ is practically absent due to its blurring. For individual directions it can still be distinguished, for example in the range $155^\circ<\theta<195^\circ$. At the same time, we can confidently trace the dependence of $\sigma_\zeta/\sigma_\xi$ on $R$. Its stratification depending on $\theta$ is less significant than for $\sigma_\eta/\sigma_\xi$, and the dependence on $R$ is obvious.

We also characterize the shape of the ellipsoid by the measure of elongation, or stretch. As such a measure, we consider the ratio 
\begin{equation}
    S = \frac{\sqrt{\sigma_\eta\sigma_\zeta}}{\sigma_\xi}
\end{equation}

As a measure of compression, we adopt the value
\begin{equation}
    C = \frac{\sqrt{\sigma_\eta\sigma_\zeta}-\sigma_\xi}{\sqrt{\sigma_\eta\sigma_\zeta}} = 1 - \frac{1}{S}
\end{equation}

As can be seen in Figs. \ref{fig:div_21}, \ref{fig:div_31}, the stretch and compression of the velocity ellipsoid depends on both $R$ and the angle $\theta$. In region of distance $R\sim11$ kpc and $165^\circ<\theta<185^\circ$ mentioned above, we note the maximum deviation of both elongation and compression from their comparatively smooth behavior in the Galactic plane. The negative sign of compression $C$ shows that the ellipsoid is not actually “compressed” like the terrestrial ellipsoid, but is elongated in the direction of the main axis $\xi$.

\begin{figure*}
    \centering
    \includegraphics[width=0.45\linewidth]{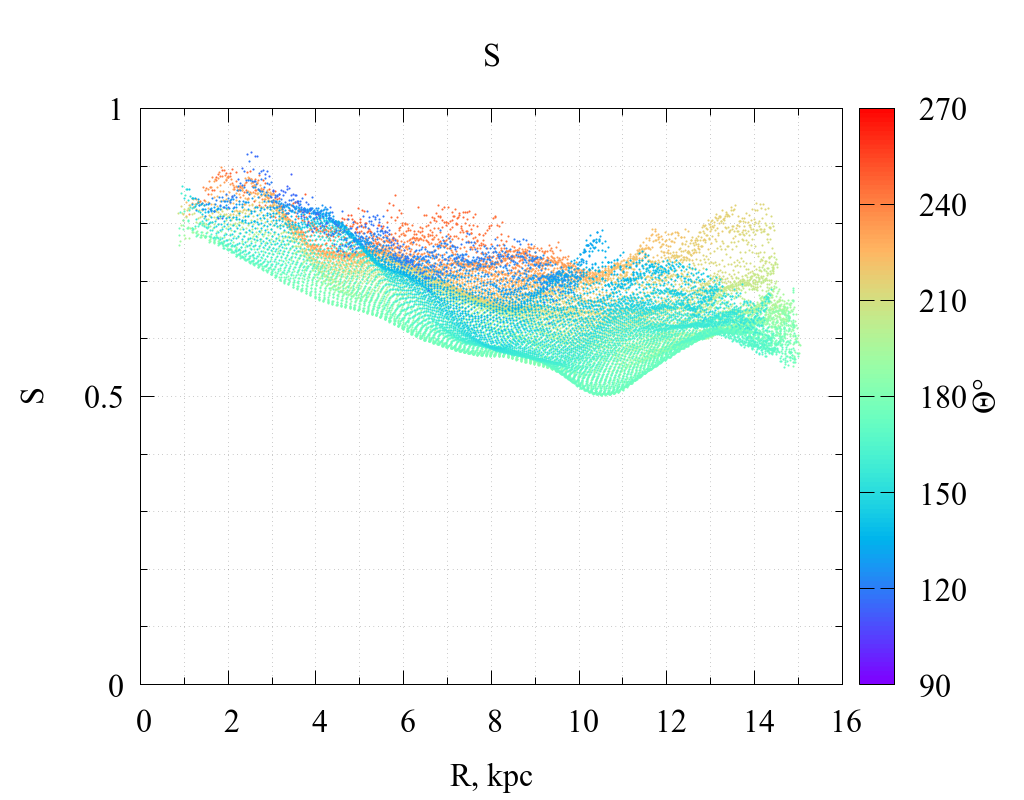}
    \includegraphics[width=0.45\linewidth]{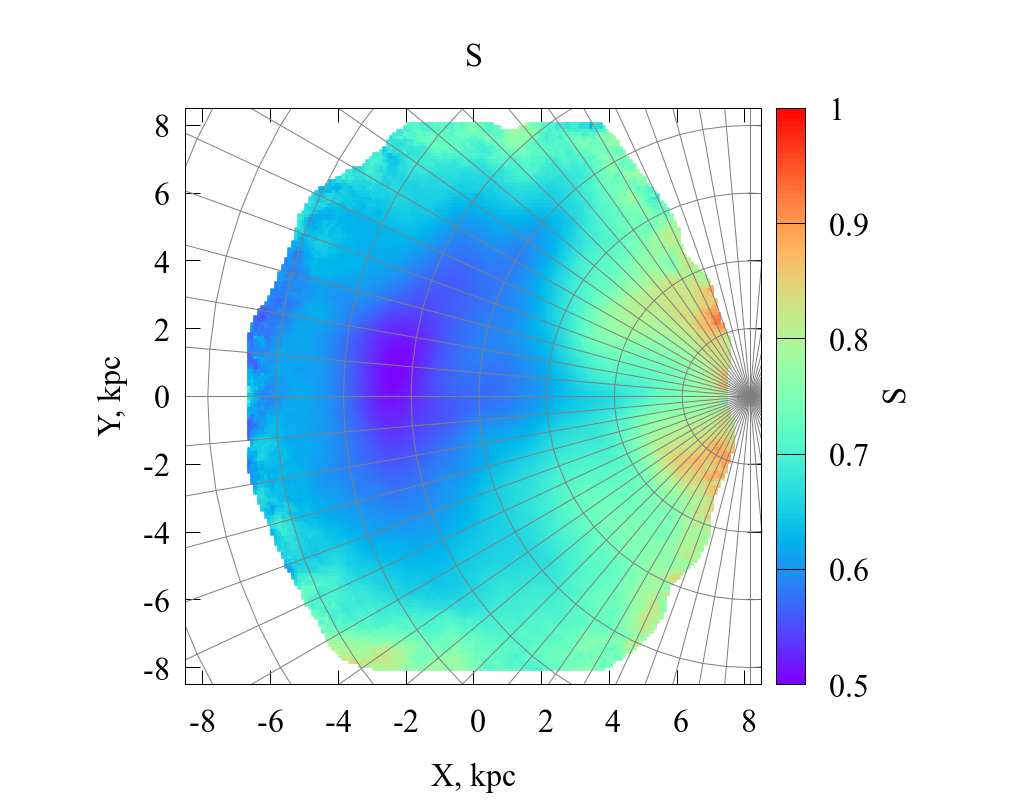}
    \caption{Left: the stretch of the velocity ellipsoid $S$ depending on the Galactocentric distance $R$, with $\theta$ shown in color. Right: the distribution map of $S$ in the rectangular Galactocentric coordinates $XY$.}
    \label{fig:S}
\end{figure*}

\begin{figure*}
    \centering
    \includegraphics[width=0.45\linewidth]{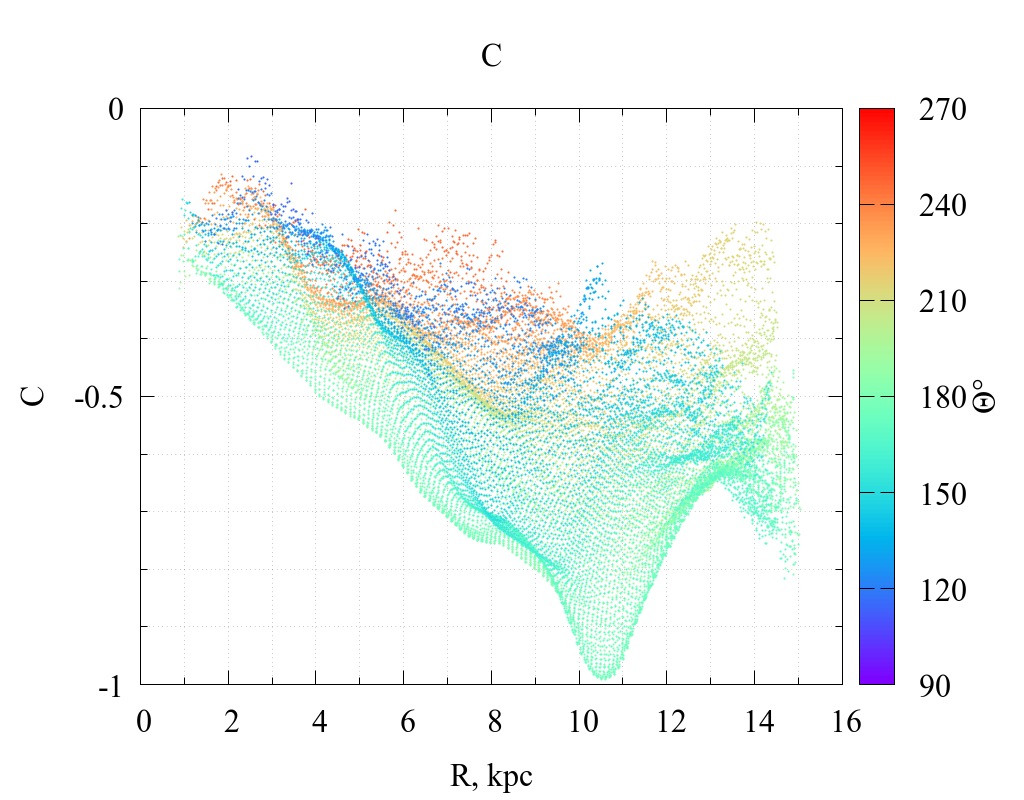}
    \includegraphics[width=0.45\linewidth]{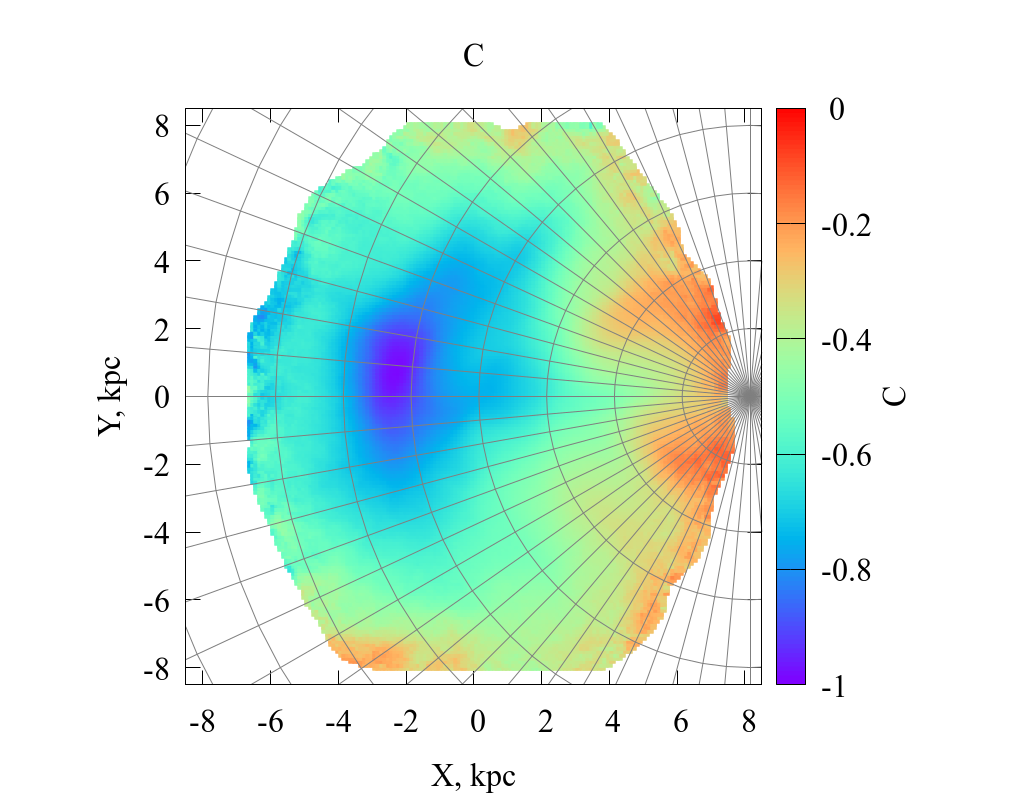}
    \caption{Left: the compression of the velocity ellipsoid $C$ depending on the Galactocentric distance $R$, with $\theta$ shown in color. Right: the distribution map of $C$ in the rectangular Galactocentric coordinates $XY$.}
    \label{fig:C}
\end{figure*}

\subsection{Comparison}
\label{sec:comparison}

\begin{figure*}
    \centering
    \includegraphics[width=0.45\linewidth]{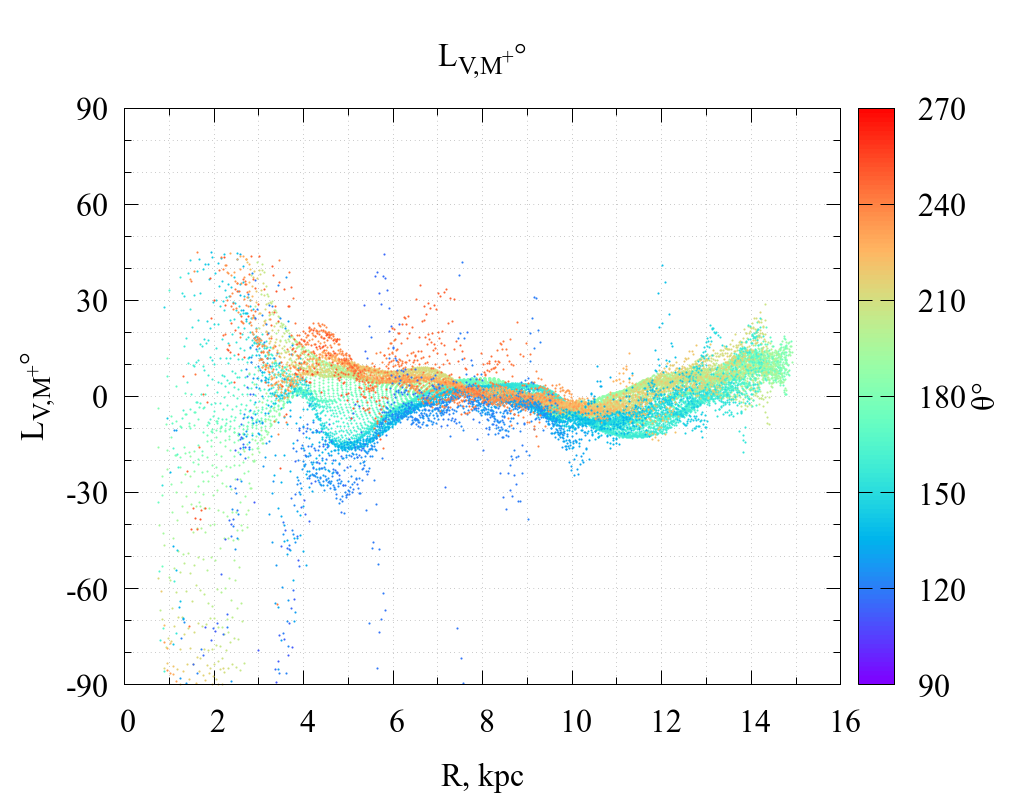}
    \includegraphics[width=0.45\linewidth]{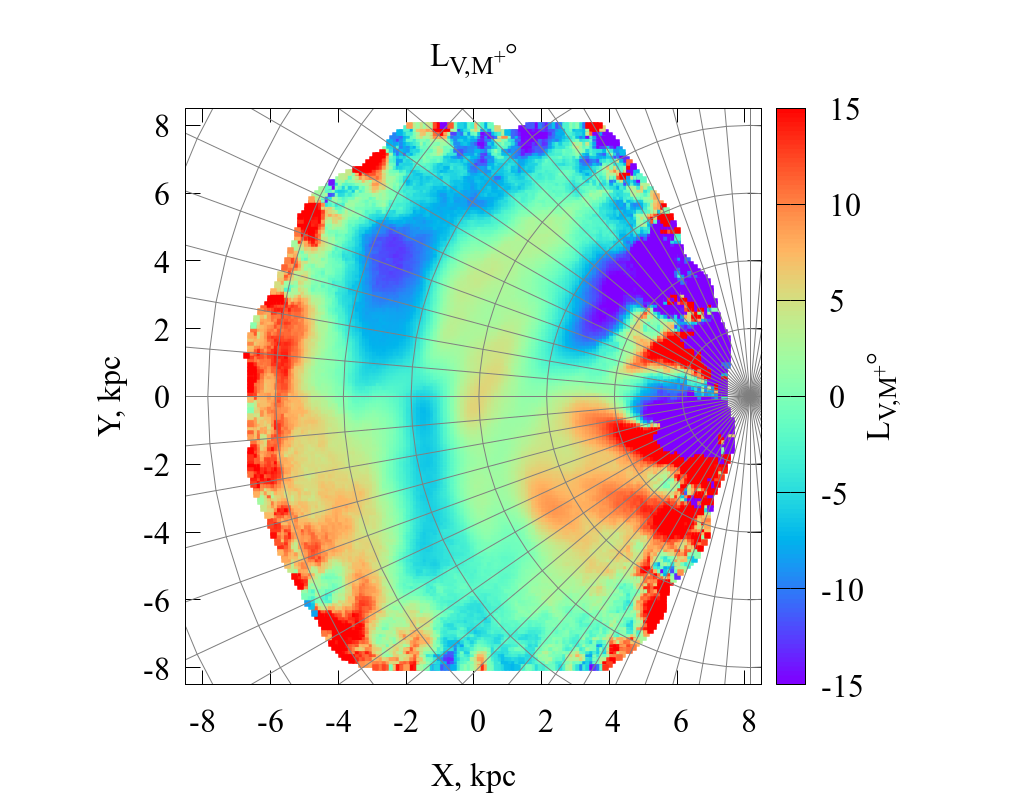}
    \caption{Left: the values of the angle $L_V$ obtained by using the components of the velocity dispersion tensor $M^+$, depending on the Galactocentric coordinate $R$ and the coordinate $\theta$, shown in color (Fig. 10 from \protect\citet{Dmytrenko2023} rebuilt). Right: the distribution map of $L_V$ in rectangular Galactocentric coordinates $XY$ (Fig. 11 from \protect\citet{Dmytrenko2023} rebuilt).}
    \label{fig:Lv_Mp}
\end{figure*}

In our previous work \citep[][]{Dmytrenko2023}, we considered how the principal axes of the deformation velocity tensors were oriented with respect to the axes of local coordinate systems. In this work, we place the main emphasis on the study of vertex deviations obtained from the components of the flat tensor $M^+$ (the $z$ dependence is not taken into account) depending on the Galactocentric coordinates. Since in the current study we obtain the vertex deviation values from the spatial velocities of stars, it will be useful to compare them with the previous results obtained \citep[$l_{xy}$ shown in Figs. 10 and 11 from][]{Dmytrenko2023}. Some of the computed parameters in \citet{Dmytrenko2023} differ from those adopted in the current study (for example, the grid step), therefore for a correct comparison, we rebuilt Figs. 10 and 11 from \citet{Dmytrenko2023} using the deformation velocity tensors ($L_{V,M^+}$ in Fig. \ref{fig:Lv_Mp}).

The behavior of $L_{V,M^+}$ and $L_\xi = L_V^\circ$ shown in Figs. \ref{fig:Lv_Mp} and \ref{fig:L1}, respectively, differs significantly from each other. Since both works used absolutely identical samples, we believe that the differences visible in the figures are caused by the differences in the plane deformation ${\bm V}_D({\bm r}_{1,2})$ and three-dimensional ${\bm V}_{\rm OBS}({\bm r}_{1,2,3})$ stellar velocity fields. The structural features visible in Figs. \ref{fig:Lv_Mp} and \ref{fig:L1} are somewhat shifted relative to each other and are less pronounced for $L_{V,M^+}$. However, a similar behavior of the $L(R)$ dependencies is also noticeable in some ranges of $R$. This is not unexpected, since both fields ${\bm V}_D({\bm r}_{1,2})$ and ${\bm V}_{\rm OBS}({\bm r}_{1,2,3})$ obviously contain information about vertex deviations. Therefore, at this point, we believe that the use of the ${\bm V}_{\rm OBS}({\bm r}_{1,2,3})$ field allows us to obtain more detailed information about its kinematic features compared to the ${\bm V}_D({\bm r}_{1,2})$ field.

\section{Summary and Conclusions}
\label{sec:conclusions}
We trace the variations in the parameters of the residual velocity ellipsoids of the \textit{Gaia} DR3 giants and subgiants belonging to thin and thick disks, depending on their position in the Galactic mid-plane. The results obtained are presented in the form of the coordinate distribution maps of the intersection points of the velocity ellipsoid semi-axes with the celestial sphere, in particular the deviations of the longitudes and latitudes of the vertices.
The distribution maps of the dispersions $\sigma_\xi, \sigma_\eta, \sigma_\zeta$ and their ratios are also given. Our results show that the use of stellar residual velocity ellipsoids for the analysis of the features of the Galaxy's kinematics allows us to obtain not only the deviation of longitudes, but also latitudes of the vertices of the stellar systems under study. In addition to the traditional conclusion that the Galaxy is not an axisymmetric system, made on the basis of variations in the angles $L_\xi$ and $L_\eta$, we found distortions of the velocity field at distances exceeding $10-12$ kpc, which manifest themselves through the deviation angles of latitudes $B_\xi$ and $B_\eta$.
These angles can be considered as kinematic signatures of the Galactic warp, obtained not through assumptions or any models, but directly from the analysis of the real velocity field. Analysis of the distribution of $L$ and $B$ at distances less than 6 kpc shows a complex kinematic picture. In this distance range, a change in the sign of the inclination of the ellipsoid planes $\xi, \eta$ relative to the Galactic plane is observed.

An analysis of the parameters characterizing the shape of the ellipsoids shows that they depend significantly on the Galactocentric distance. The strongest dependence is observed at distances $R<6$ kpc, and then it becomes much smoother and at 14 kpc $\sigma_\xi, \sigma_\eta, \sigma_\zeta$ have approximate values of 26.5, 17.2, 14.7 \kms, respectively. Their dependence on $\theta$ is also observed. However, this azimuthal dependence, displayed in the figures as a stratification, is insignificant. The strongest stratification is observed for $\sigma_\eta$. At the distance of the Sun, it reaches a value of approximately 20 \kms, and at a distance of 14 kpc it becomes less than 10 \kms. The ratio of the velocity ellipsoid semi-axes lengths indicates a special region in the direction of the Galactic anticenter at a distance of approximately 11 kpc. In the same place of the Galactic plane, the elongation of the ellipsoids differs significantly (by $1.5-2$ times) from that in the rest of the plane. We plan to interpret these observed facts in our following works.

Based on the comparison of the results of our two works, we believe that the use of a three-dimensional velocity field allows us to obtain more detailed information about the kinematic features of our Galaxy compared to the use of the ${\bm V}_D({\bm r}_{1,2})$ field.

\section{Acknowledgements}
\label{sec:acknowledgements}

The authors are grateful to the Armed Forces of Ukraine for ensuring the security of our country and the opportunity for scientists to conduct research. 

This work is supported by the National Research Foundation of Ukraine, Project No. 2023.03/0188, and the Ministry of Science of Ukraine.

This work has made use of data from the European Space Agency (ESA) mission {\it Gaia} (\url{https://www.cosmos.esa.int/gaia}), processed by the {\it Gaia} Data Processing and Analysis Consortium (DPAC, \url{https://www.cosmos.esa.int/web/gaia/dpac/consortium}). Funding for the DPAC has been provided by national institutions, in particular the institutions participating in the {\it Gaia} Multilateral Agreement.

\section*{Data availability}
\addcontentsline{toc}{section}{Data availability}
The used catalogue data is available in a standardised format for readers via the CDS (https://cds.u-strasbg.fr).
The software code used in this paper and the parameters characterizing the shapes of velocity ellipsoids can be made available on personal request by e-mail: \href{mailto:astronom.karazin007@gmail.com}  {astronom.karazin007@gmail.com} or \href{mailto:akhmetovvs@gmail.com} {akhmetovvs@gmail.com}.

\bibliographystyle{mnras}
\bibliography{references}

\begin{thebibliography}{}
\makeatletter
\relax
\def\mn@urlcharsother{\let\do\@makeother \do\$\do\&\do\#\do\^\do\_\do\%\do\~}
\def\mn@doi{\begingroup\mn@urlcharsother \@ifnextchar [ {\mn@doi@} {\mn@doi@[]}}
\def\mn@doi@[#1]#2{\def\@tempa{#1}\ifx\@tempa\@empty \href {http://dx.doi.org/#2} {doi:#2}\else \href {http://dx.doi.org/#2} {#1}\fi \endgroup}
\def\mn@eprint#1#2{\mn@eprint@#1:#2::\@nil}
\def\mn@eprint@arXiv#1{\href {http://arxiv.org/abs/#1} {{\tt arXiv:#1}}}
\def\mn@eprint@dblp#1{\href {http://dblp.uni-trier.de/rec/bibtex/#1.xml} {dblp:#1}}
\def\mn@eprint@#1:#2:#3:#4\@nil{\def\@tempa {#1}\def\@tempb {#2}\def\@tempc {#3}\ifx \@tempc \@empty \let \@tempc \@tempb \let \@tempb \@tempa \fi \ifx \@tempb \@empty \def\@tempb {arXiv}\fi \@ifundefined {mn@eprint@\@tempb}{\@tempb:\@tempc}{\expandafter \expandafter \csname mn@eprint@\@tempb\endcsname \expandafter{\@tempc}}}

\bibitem[\protect\citeauthoryear{{Akhmetov}, {Bucciarelli}, {Crosta}, {Lattanzi}, {Spagna}, {Re Fiorentin}  \& {Bannikova}}{{Akhmetov} et~al.}{2024}]{Akhmetov2024}
{Akhmetov} V.~S.,  {Bucciarelli} B.,  {Crosta} M.,  {Lattanzi} M.~G.,  {Spagna} A.,  {Re Fiorentin} P.,   {Bannikova} E.~Y.,  2024, \mn@doi [\mnras] {10.1093/mnras/stae772}, \href {https://ui.adsabs.harvard.edu/abs/2024MNRAS.530..710A} {530, 710}

\bibitem[\protect\citeauthoryear{{Amendt} \& {Cuddeford}}{{Amendt} \& {Cuddeford}}{1991}]{Amendt1991}
{Amendt} P.,  {Cuddeford} P.,  1991, \mn@doi [\apj] {10.1086/169672}, \href {https://ui.adsabs.harvard.edu/abs/1991ApJ...368...79A} {368, 79}

\bibitem[\protect\citeauthoryear{{Antoja} et~al.,}{{Antoja} et~al.}{2018}]{Antoja2018}
{Antoja} T.,  et~al., 2018, \mn@doi [\nat] {10.1038/s41586-018-0510-7}, \href {https://ui.adsabs.harvard.edu/abs/2018Natur.561..360A} {561, 360}

\bibitem[\protect\citeauthoryear{{Bailer-Jones}, {Rybizki}, {Fouesneau}, {Demleitner}  \& {Andrae}}{{Bailer-Jones} et~al.}{2021}]{Bailer-Jones2021}
{Bailer-Jones} C.~A.~L.,  {Rybizki} J.,  {Fouesneau} M.,  {Demleitner} M.,   {Andrae} R.,  2021, \mn@doi [\aj] {10.3847/1538-3881/abd806}, \href {https://ui.adsabs.harvard.edu/abs/2021AJ....161..147B} {161, 147}

\bibitem[\protect\citeauthoryear{{Binney} \& {Tremaine}}{{Binney} \& {Tremaine}}{2008}]{Binney2008}
{Binney} J.,  {Tremaine} S.,  2008, {Galactic Dynamics: Second Edition}.
Princeton University Press

\bibitem[\protect\citeauthoryear{{Cantat-Gaudin} \& {Brandt}}{{Cantat-Gaudin} \& {Brandt}}{2021}]{Cantat-Gaudin2021}
{Cantat-Gaudin} T.,  {Brandt} T.~D.,  2021, \mn@doi [\aap] {10.1051/0004-6361/202140807}, \href {https://ui.adsabs.harvard.edu/abs/2021A&A...649A.124C} {649, A124}

\bibitem[\protect\citeauthoryear{{Dehnen}}{{Dehnen}}{2000}]{Dehnen2000}
{Dehnen} W.,  2000, \mn@doi [\aj] {10.1086/301226}, \href {https://ui.adsabs.harvard.edu/abs/2000AJ....119..800D} {119, 800}

\bibitem[\protect\citeauthoryear{{Dmytrenko}, {Fedorov}, {Akhmetov}, {Velichko}  \& {Denyshchenko}}{{Dmytrenko} et~al.}{2023}]{Dmytrenko2023}
{Dmytrenko} A.~M.,  {Fedorov} P.~N.,  {Akhmetov} V.~S.,  {Velichko} A.~B.,   {Denyshchenko} S.~I.,  2023, \mn@doi [\mnras] {10.1093/mnras/stad823}, \href {https://ui.adsabs.harvard.edu/abs/2023MNRAS.521.4247D} {521, 4247}

\bibitem[\protect\citeauthoryear{{Fedorov}, {Akhmetov}, {Velichko}, {Dmytrenko}  \& {Denischenko}}{{Fedorov} et~al.}{2021}]{Fedorov2021}
{Fedorov} P.~N.,  {Akhmetov} V.~S.,  {Velichko} A.~B.,  {Dmytrenko} A.~M.,   {Denischenko} S.~I.,  2021, \mn@doi [\mnras] {10.1093/mnras/stab2821}, \href {https://ui.adsabs.harvard.edu/abs/2021MNRAS.508.3055F} {508, 3055}

\bibitem[\protect\citeauthoryear{{Fedorov}, {Akhmetov}, {Velichko}, {Dmytrenko}  \& {Denyshchenko}}{{Fedorov} et~al.}{2023}]{Fedorov2023}
{Fedorov} P.~N.,  {Akhmetov} V.~S.,  {Velichko} A.~B.,  {Dmytrenko} A.~M.,   {Denyshchenko} S.~I.,  2023, \mn@doi [\mnras] {10.1093/mnras/stac3218}, \href {https://ui.adsabs.harvard.edu/abs/2023MNRAS.518.2761F} {518, 2761}

\bibitem[\protect\citeauthoryear{{GRAVITY Collaboration} et~al.,}{{GRAVITY Collaboration} et~al.}{2021}]{Abuter2021}
{GRAVITY Collaboration} et~al., 2021, \mn@doi [\aap] {10.1051/0004-6361/202040208}, \href {https://ui.adsabs.harvard.edu/abs/2021A&A...647A..59G} {647, A59}

\bibitem[\protect\citeauthoryear{{Gaia Collaboration} et~al.,}{{Gaia Collaboration} et~al.}{2016}]{Prusti2016}
{Gaia Collaboration} et~al., 2016, \mn@doi [\aap] {10.1051/0004-6361/201629272}, \href {https://ui.adsabs.harvard.edu/abs/2016A&A...595A...1G} {595, A1}

\bibitem[\protect\citeauthoryear{{Gaia Collaboration} et~al.,}{{Gaia Collaboration} et~al.}{2018}]{Helmi2018}
{Gaia Collaboration} et~al., 2018, \mn@doi [\aap] {10.1051/0004-6361/201832698}, \href {https://ui.adsabs.harvard.edu/abs/2018A&A...616A..12G} {616, A12}

\bibitem[\protect\citeauthoryear{{Gaia Collaboration} et~al.,}{{Gaia Collaboration} et~al.}{2021}]{Antoja2021}
{Gaia Collaboration} et~al., 2021, \mn@doi [\aap] {10.1051/0004-6361/202039714}, \href {https://ui.adsabs.harvard.edu/abs/2021A&A...649A...8G} {649, A8}

\bibitem[\protect\citeauthoryear{{Gaia Collaboration} et~al.,}{{Gaia Collaboration} et~al.}{2023a}]{Vallenari2023}
{Gaia Collaboration} et~al., 2023a, \mn@doi [\aap] {10.1051/0004-6361/202243940}, \href {https://ui.adsabs.harvard.edu/abs/2023A&A...674A...1G} {674, A1}

\bibitem[\protect\citeauthoryear{{Gaia Collaboration} et~al.,}{{Gaia Collaboration} et~al.}{2023b}]{Drimmel2023}
{Gaia Collaboration} et~al., 2023b, \mn@doi [\aap] {10.1051/0004-6361/202243797}, \href {https://ui.adsabs.harvard.edu/abs/2023A&A...674A..37G} {674, A37}

\bibitem[\protect\citeauthoryear{{Kuijken} \& {Gilmore}}{{Kuijken} \& {Gilmore}}{1991}]{Kuijken1991}
{Kuijken} K.,  {Gilmore} G.,  1991, \mn@doi [\apjl] {10.1086/185920}, \href {https://ui.adsabs.harvard.edu/abs/1991ApJ...367L...9K} {367, L9}

\bibitem[\protect\citeauthoryear{{Lindegren} et~al.,}{{Lindegren} et~al.}{2018}]{Lindegren2018}
{Lindegren} L.,  et~al., 2018, \mn@doi [\aap] {10.1051/0004-6361/201832727}, \href {https://ui.adsabs.harvard.edu/abs/2018A&A...616A...2L} {616, A2}

\bibitem[\protect\citeauthoryear{{Lindegren} et~al.,}{{Lindegren} et~al.}{2021}]{Lindegren2021}
{Lindegren} L.,  et~al., 2021, \mn@doi [\aap] {10.1051/0004-6361/202039653}, \href {https://ui.adsabs.harvard.edu/abs/2021A&A...649A...4L} {649, A4}

\bibitem[\protect\citeauthoryear{{Minchev} \& {Famaey}}{{Minchev} \& {Famaey}}{2010}]{Minchev2010}
{Minchev} I.,  {Famaey} B.,  2010, \mn@doi [\apj] {10.1088/0004-637X/722/1/112}, \href {https://ui.adsabs.harvard.edu/abs/2010ApJ...722..112M} {722, 112}

\bibitem[\protect\citeauthoryear{{Parenago}}{{Parenago}}{1951}]{Parenago1951}
{Parenago} P.~P.,  1951, Trudy Gosudarstvennogo Astronomicheskogo Instituta, \href {https://ui.adsabs.harvard.edu/abs/1951TrSht..20...26P} {20, 26}

\bibitem[\protect\citeauthoryear{{Perryman}, {Lindegren}, {Kovalevsky}  \& {et al.}}{{Perryman} et~al.}{1997}]{Perryman1997}
{Perryman} M. A.~C.,  {Lindegren} L.,  {Kovalevsky} J.,   {et al.} 1997, \mn@doi [A\&A] {10.1051/0004-6361:19971013}, \href {https://ui.adsabs.harvard.edu/abs/1997A%26A...323L..49P} {323, L49}

\bibitem[\protect\citeauthoryear{{Rootsm\"{a}e}}{{Rootsm\"{a}e}}{1959}]{Rootsmae1959}
{Rootsm\"{a}e} T.,  1959, Tartu Riiklik \"{U}likool toimetsced. Astronoomia, f\"{u}\"{u}sika ja keemia-alaseid t\"{o}id. [in Russian], 74, 3

\bibitem[\protect\citeauthoryear{{Saha}, {Pfenniger}  \& {Taam}}{{Saha} et~al.}{2013}]{Saha2013}
{Saha} K.,  {Pfenniger} D.,   {Taam} R.~E.,  2013, \mn@doi [\apj] {10.1088/0004-637X/764/2/123}, \href {https://ui.adsabs.harvard.edu/abs/2013ApJ...764..123S} {764, 123}

\bibitem[\protect\citeauthoryear{{Smith}, {Whiteoak}  \& {Evans}}{{Smith} et~al.}{2012}]{Smith2012}
{Smith} M.~C.,  {Whiteoak} S.~H.,   {Evans} N.~W.,  2012, \mn@doi [\apj] {10.1088/0004-637X/746/2/181}, \href {https://ui.adsabs.harvard.edu/abs/2012ApJ...746..181S} {746, 181}

\bibitem[\protect\citeauthoryear{{Tarapov}}{{Tarapov}}{2002}]{Tarapov2002}
{Tarapov} I.~E.,  2002, Continuum Mechanics: General Laws of Kinematics and Dynamics.
~ Vol. 2, Gould Pages, Kharkiv, Ukraine [In Russian]

\bibitem[\protect\citeauthoryear{{Vorobyov} \& {Theis}}{{Vorobyov} \& {Theis}}{2008}]{Vorobyov2008}
{Vorobyov} E.~I.,  {Theis} C.,  2008, \mn@doi [\mnras] {10.1111/j.1365-2966.2007.12476.x}, \href {https://ui.adsabs.harvard.edu/abs/2008MNRAS.383..817V} {383, 817}

\makeatother
\end{thebibliography}

\begin{appendix}
\section{Uncertainties of \texorpdfstring{$L$ and $B$}{L and B}}
\label{sec:sigma_LB}

\begin{figure*}
    \centering
    \includegraphics[width=0.45\linewidth]{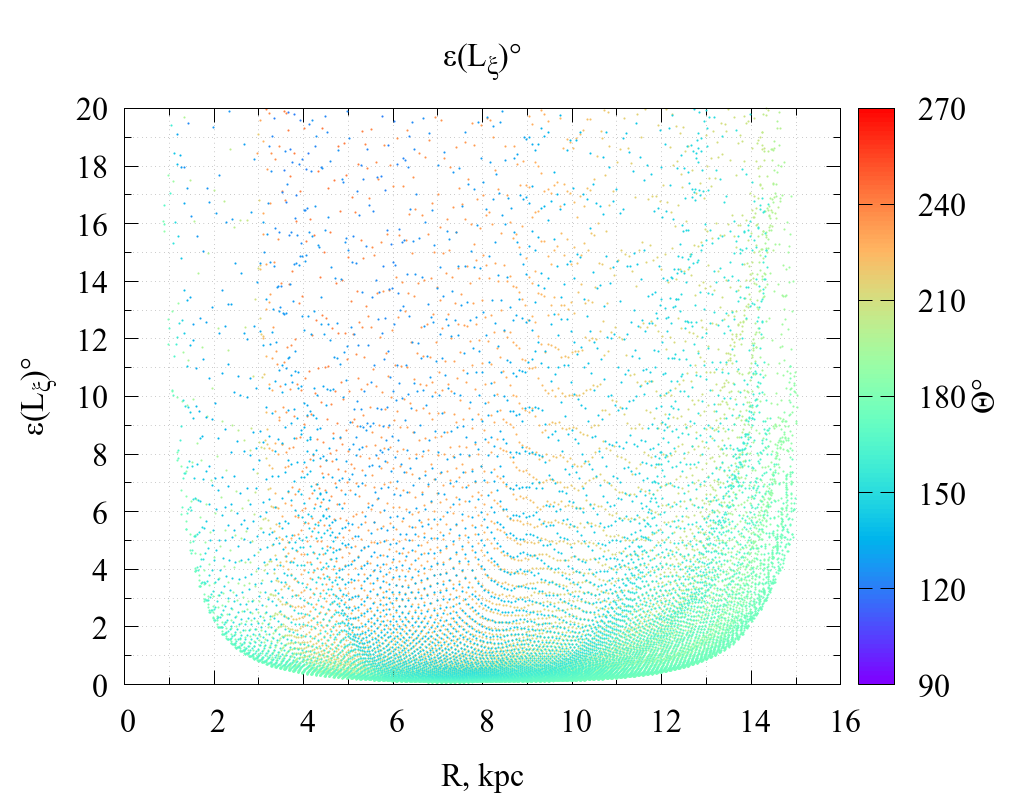}
    \includegraphics[width=0.45\linewidth]{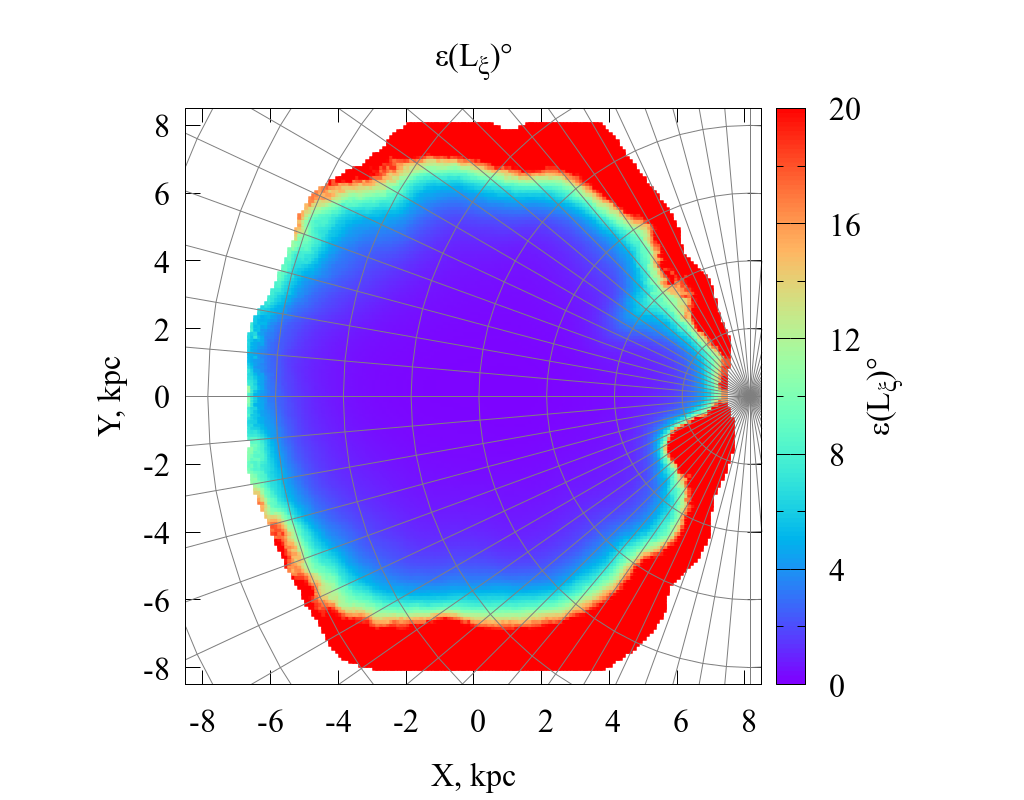}
    \caption{Left: uncertainties of the angle $L_\xi$ depending on the Galactocentric distance $R$, with $\theta$ shown in color. Right: the distribution map of $L_\xi$ uncertainties on the rectangular Galactocentric coordinates $XY$.}
    \label{fig:eL1}
\end{figure*}

    \begin{figure*}
    \centering
    \includegraphics[width=0.45\linewidth]{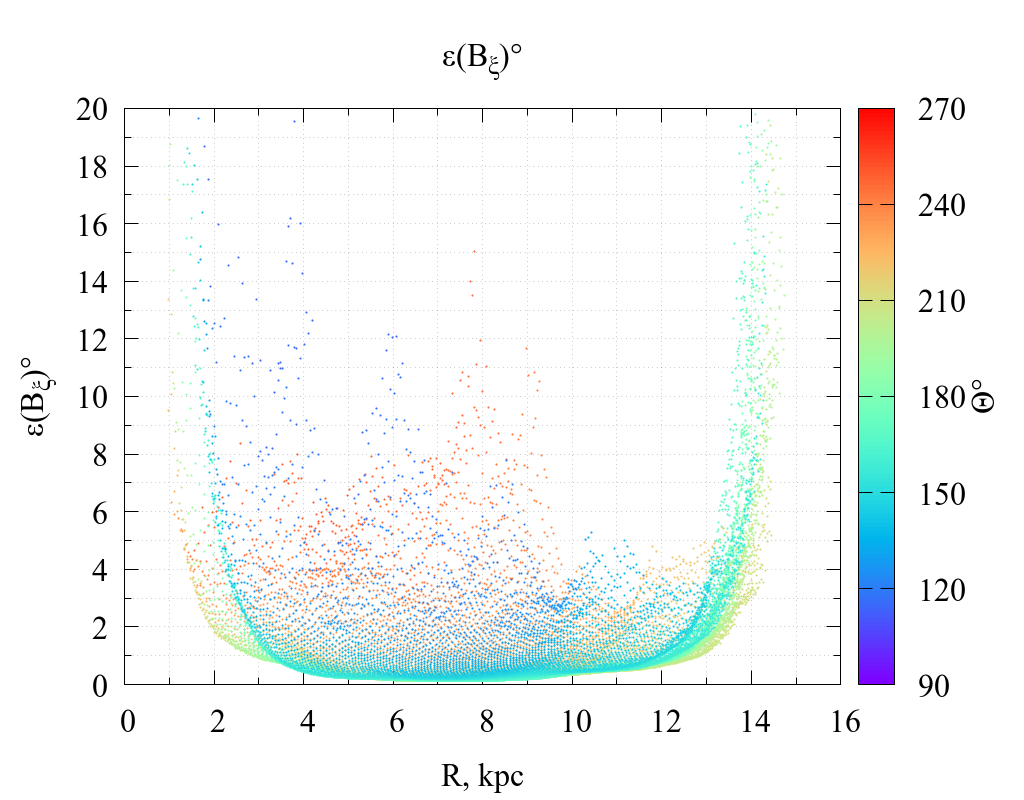}
    \includegraphics[width=0.45\linewidth]{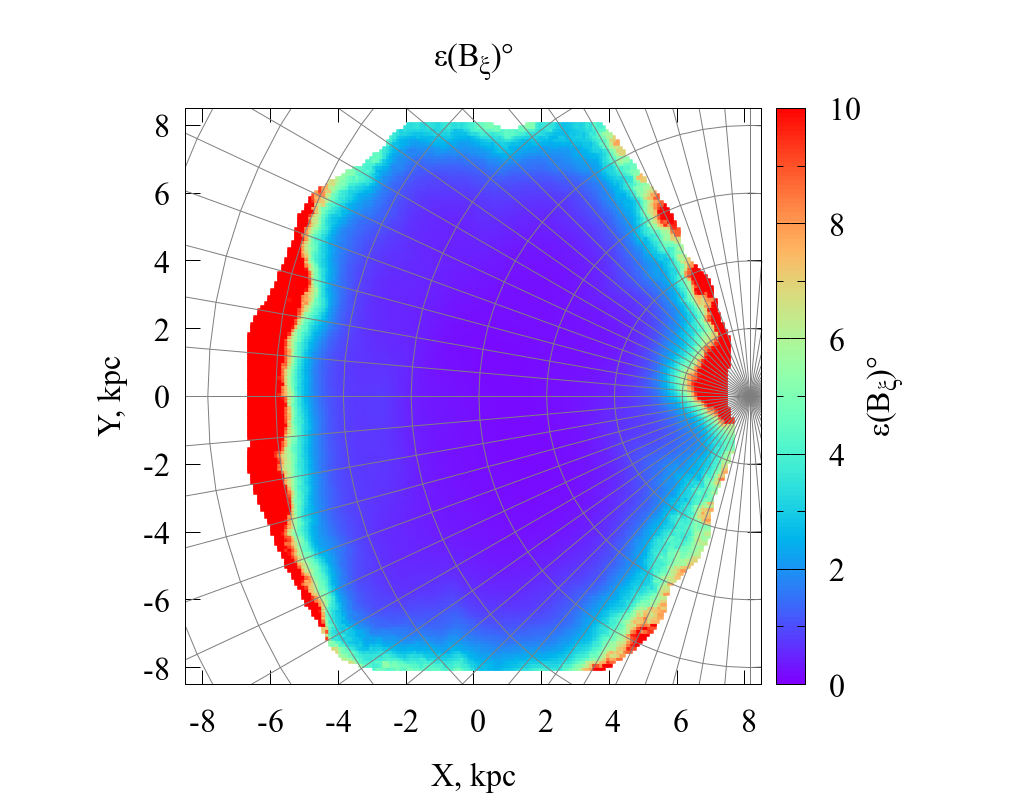}
    \caption{Left: uncertainties of the angle $B_\xi$ depending on the Galactocentric distance $R$, with $\theta$ shown in color. Right: the distribution map of $B_\xi$ uncertainties on the rectangular Galactocentric coordinates $XY$.}
    \label{fig:eB1}
\end{figure*}

\begin{figure*}
    \centering
    \includegraphics[width=0.45\linewidth]{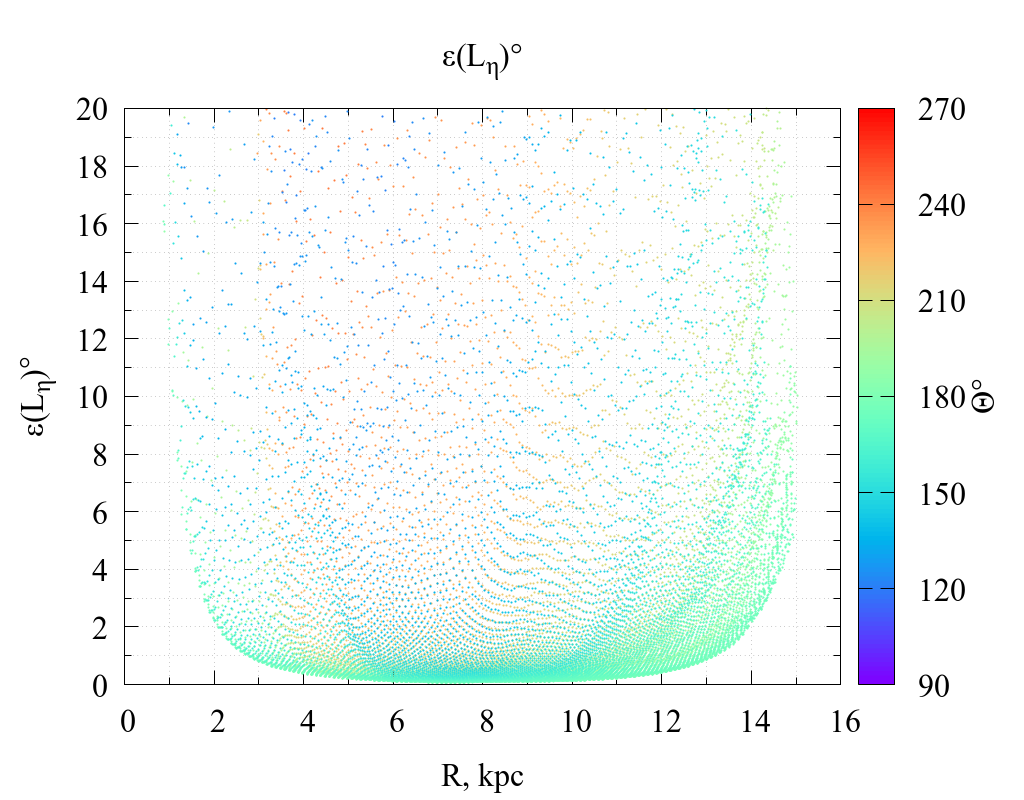}
    \includegraphics[width=0.45\linewidth]{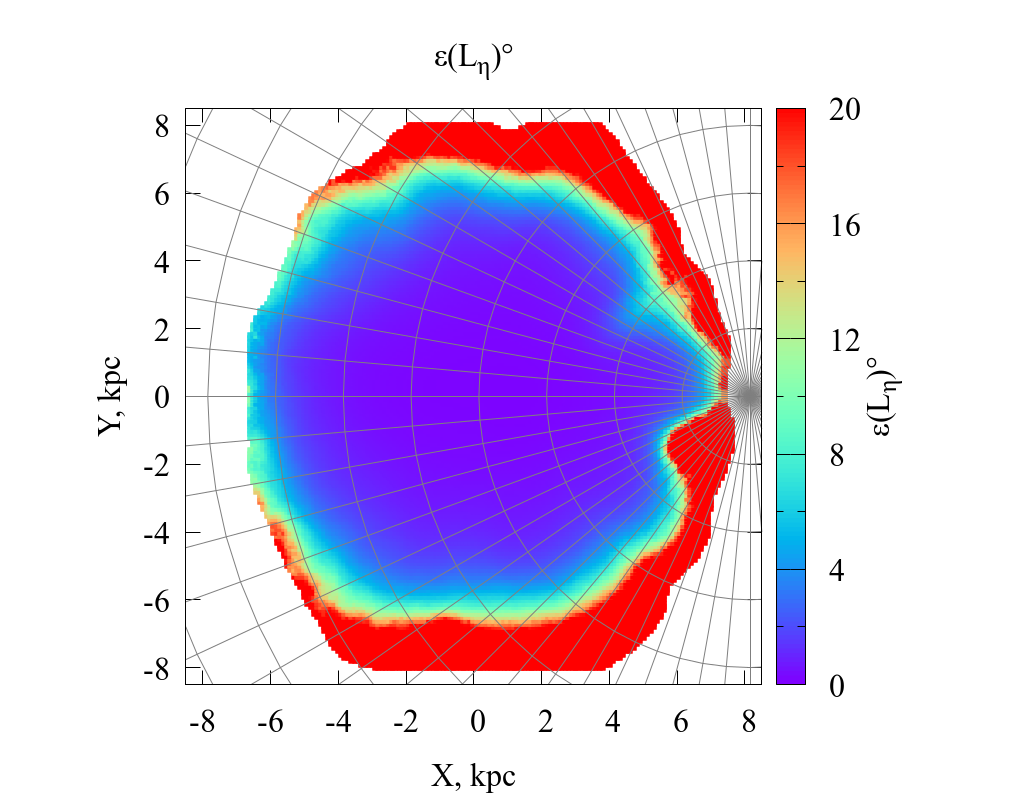}
    \caption{Left: uncertainties of the angle $L_\eta$ depending on the Galactocentric distance $R$, with $\theta$ shown in color. Right: the distribution map of $L_\eta$ uncertainties on the rectangular Galactocentric coordinates $XY$.}
    \label{fig:eL2}
\end{figure*}

\begin{figure*}
    \centering
    \includegraphics[width=0.45\linewidth]{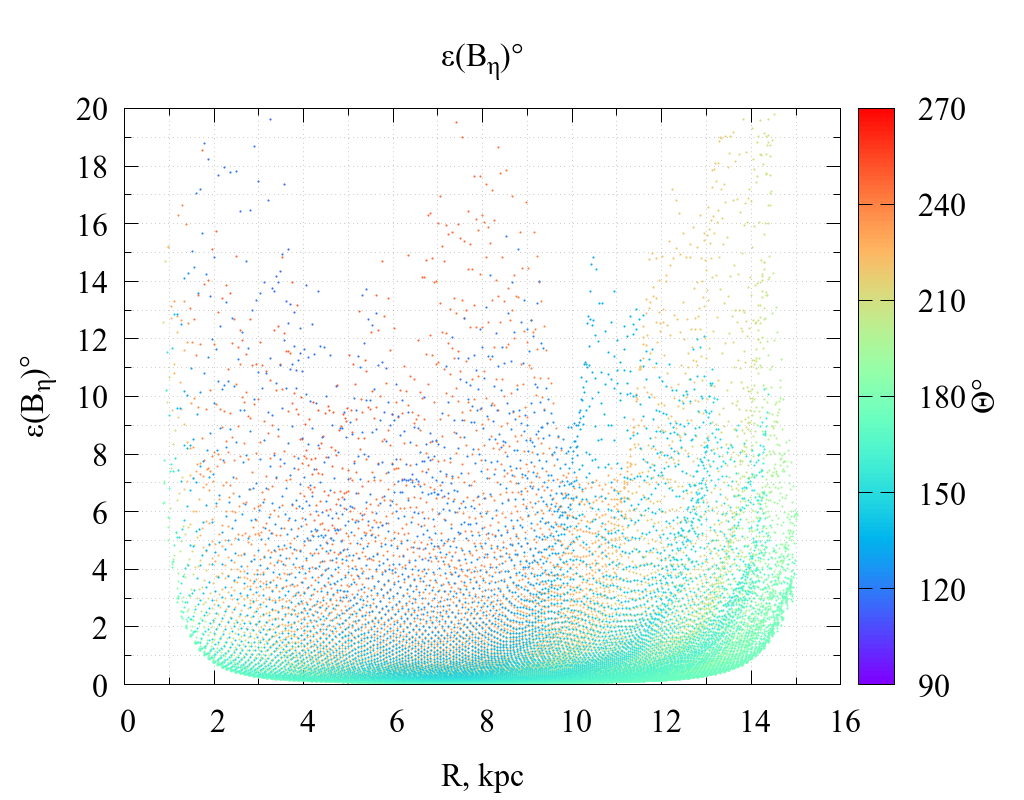}
    \includegraphics[width=0.45\linewidth]{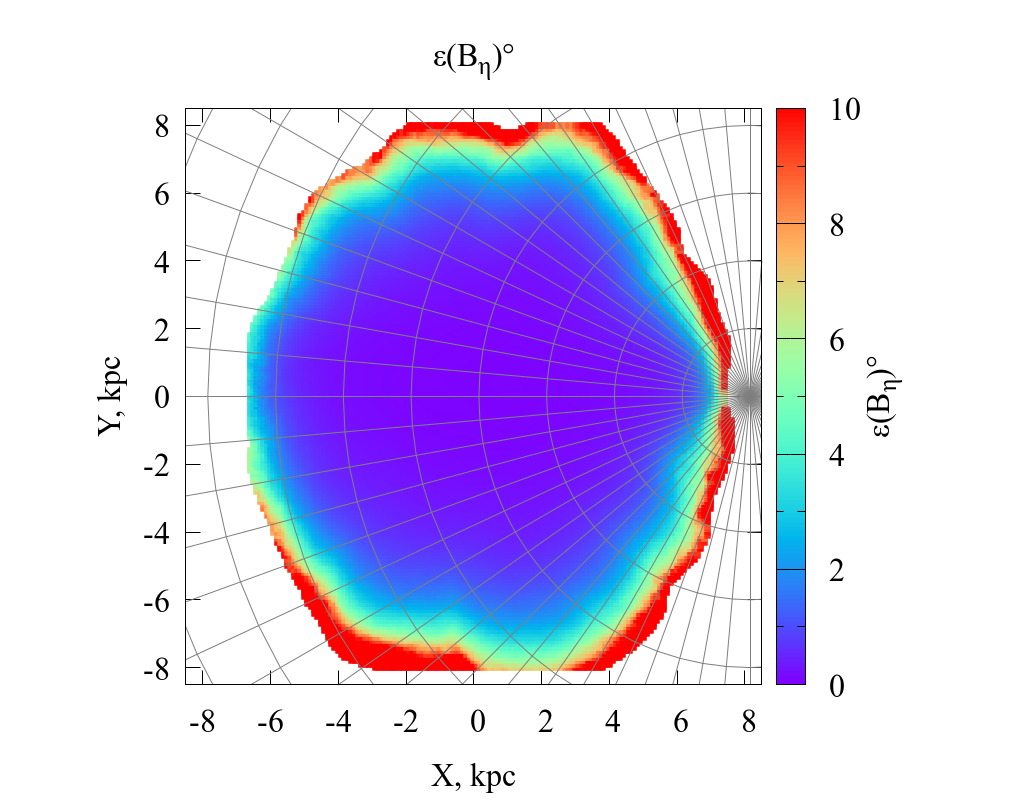}
    \caption{Left: uncertainties of the angle $B_\eta$ depending on the Galactocentric distance $R$, with $\theta$ shown in color. Right: the distribution map of $B_\eta$ uncertainties on the rectangular Galactocentric coordinates $XY$.}
    \label{fig:eB2}
\end{figure*}

\begin{figure*}
    \centering
    \includegraphics[width=0.45\linewidth]{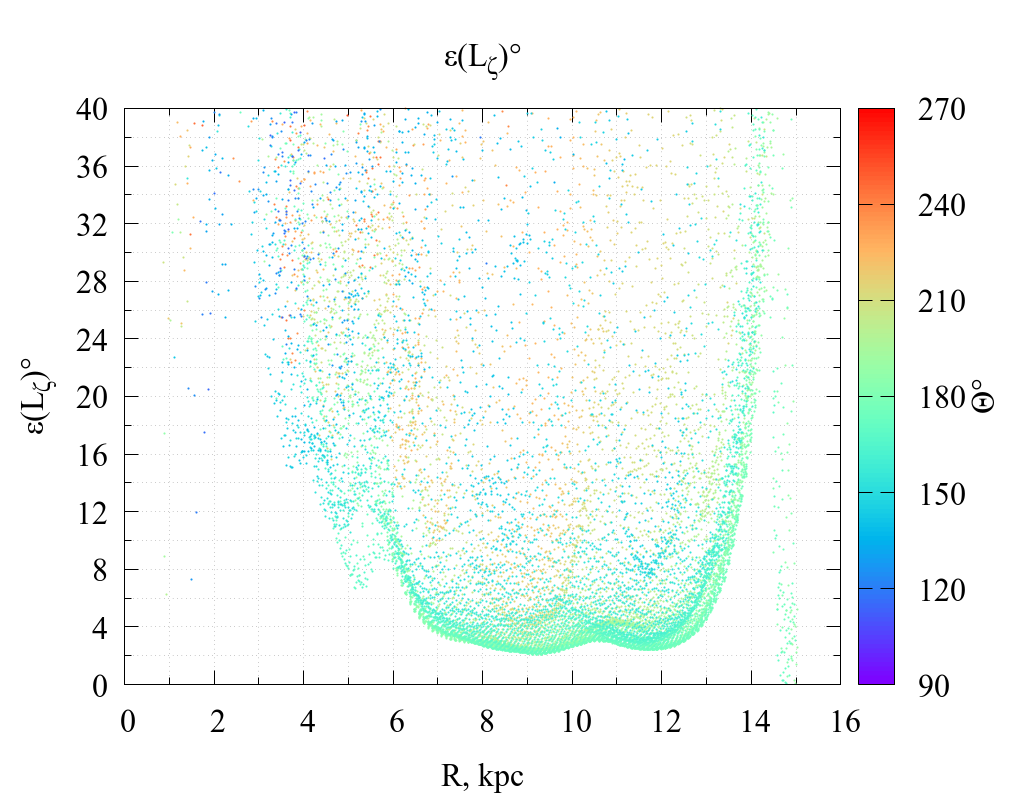}
    \includegraphics[width=0.45\linewidth]{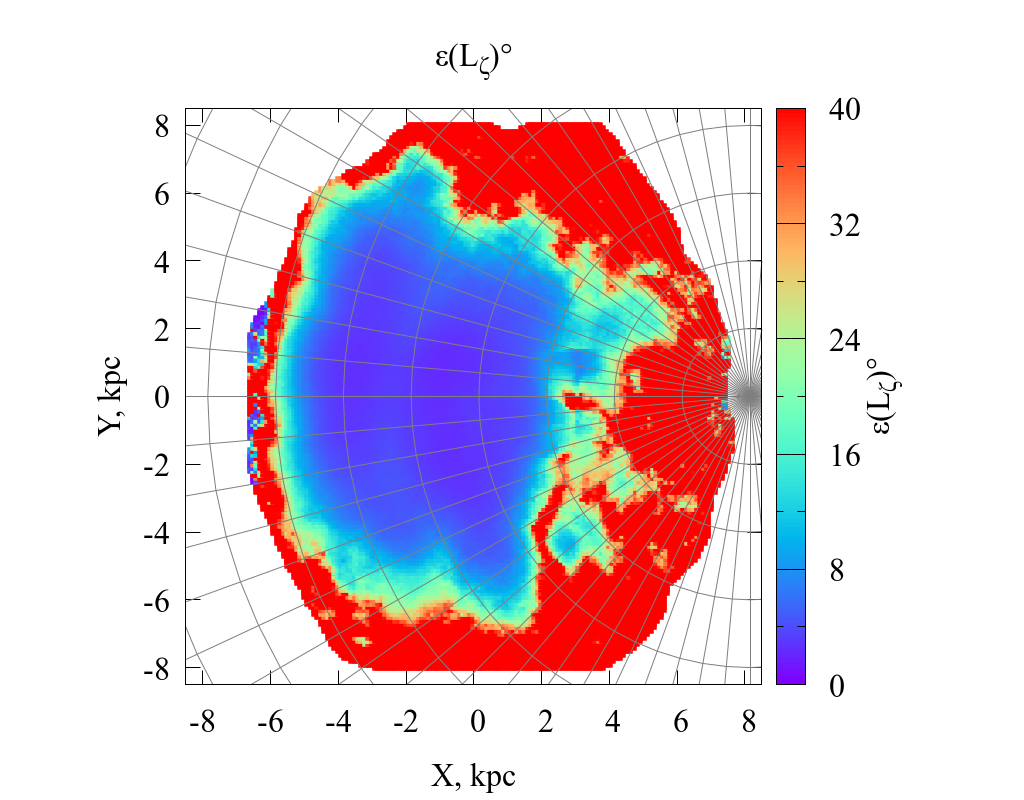}
    \caption{Left: uncertainties of the angle $L_\zeta$ depending on the Galactocentric distance $R$, with $\theta$ shown in color. Right: the distribution map of $L_\zeta$ uncertainties on the rectangular Galactocentric coordinates $XY$.}
    \label{fig:eL3}
\end{figure*}

\begin{figure*}
    \centering
    \includegraphics[width=0.45\linewidth]{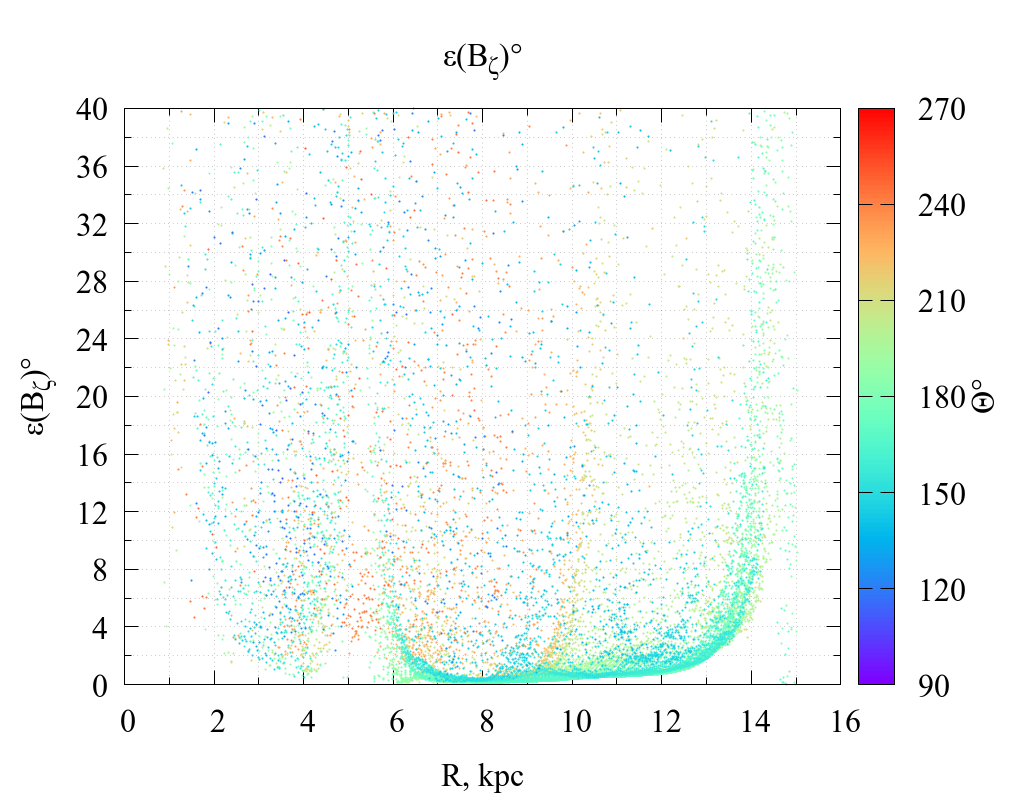}
    \includegraphics[width=0.45\linewidth]{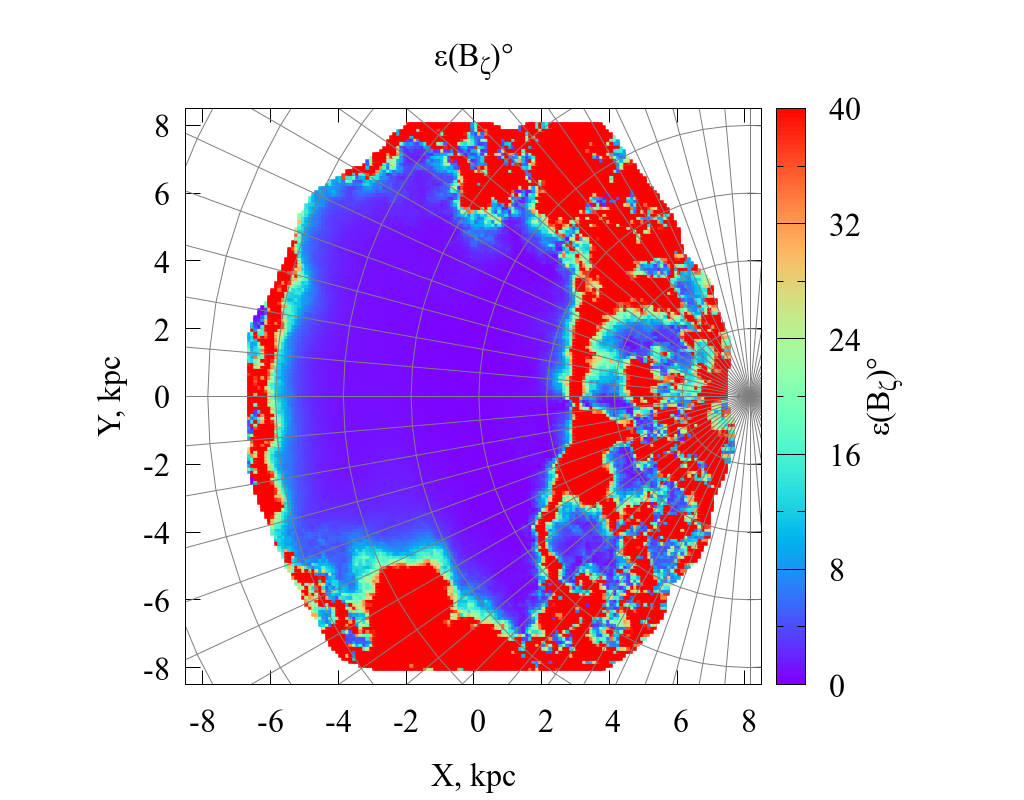}
    \caption{Left: uncertainties of the angle $B_\zeta$ depending on the Galactocentric distance $R$, with $\theta$ shown in color. Right: the distribution map of $B_\zeta$ uncertainties on the rectangular Galactocentric coordinates $XY$.}
    \label{fig:eB3}
\end{figure*}

\section{Uncertainties of \texorpdfstring{$\sigma_\xi, \sigma_\eta, \sigma_\zeta$}{sigma xi, sigma eta, sigma zeta}}
\label{sec:sigma_123}

\begin{figure*}
    \centering
    \includegraphics[width=0.45\linewidth]{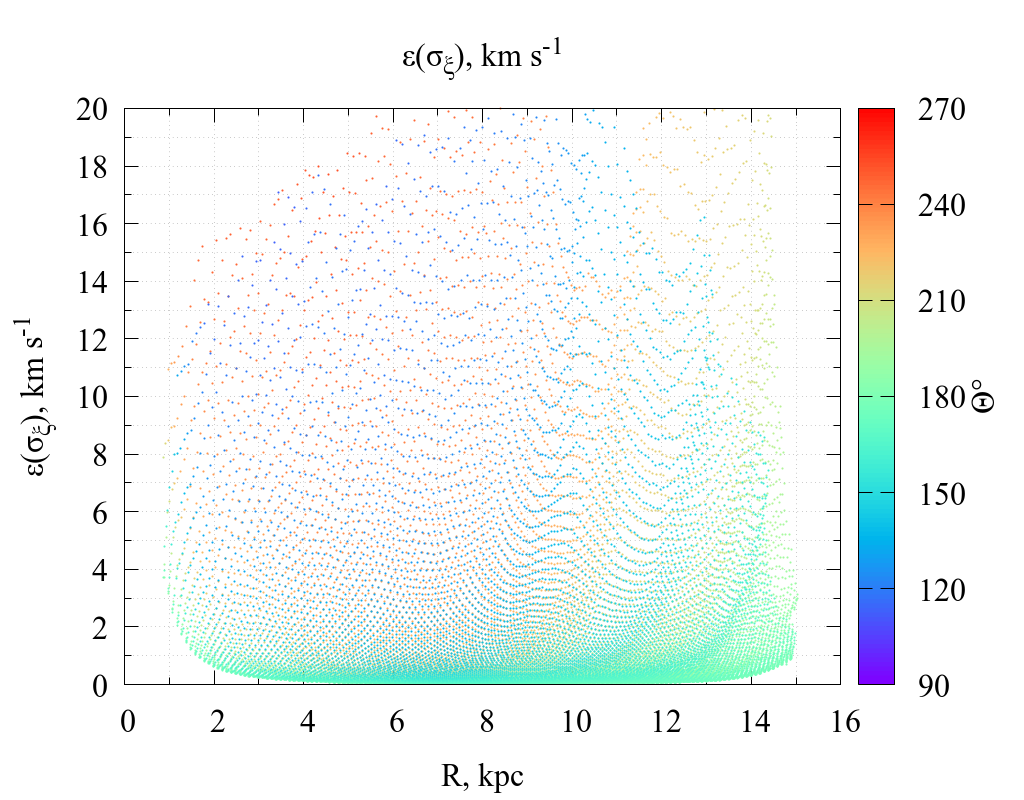}
    \includegraphics[width=0.45\linewidth]{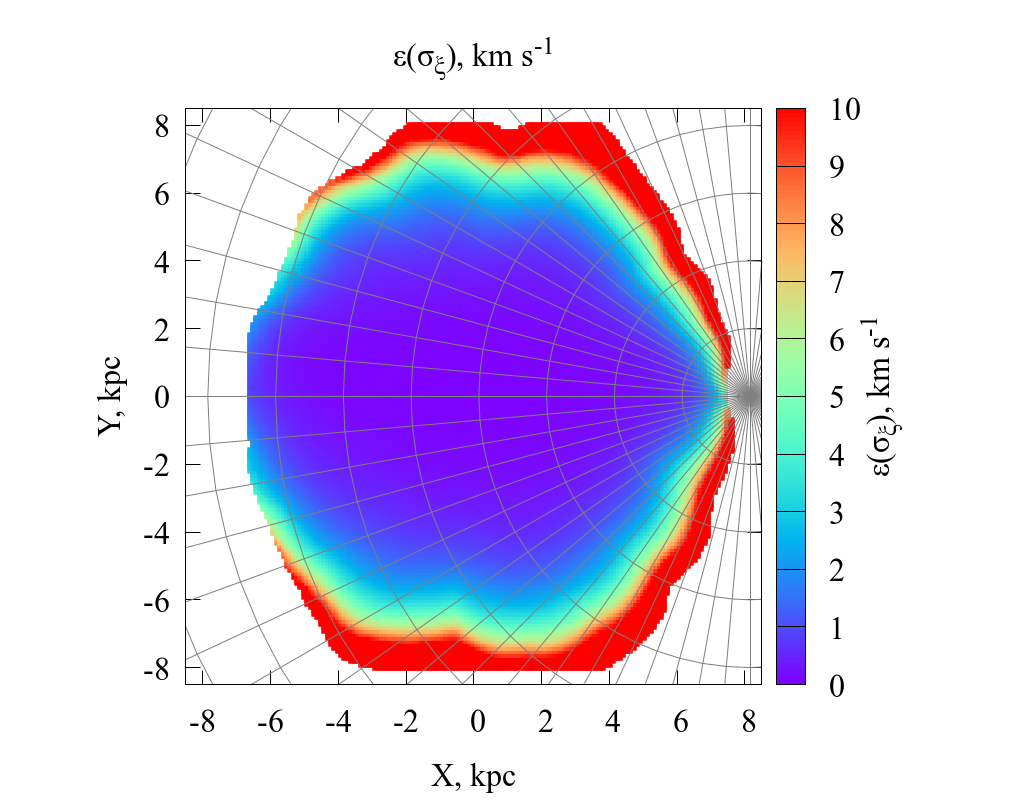}
    \caption{Left: uncertainties of the velocity ellipsoid semi-axis lengths $\varepsilon(\sigma_\xi)$ depending on the Galactocentric distance $R$, with $\theta$ shown in color. Right: the distribution map of $\varepsilon(\sigma_\xi)$ on the rectangular Galactocentric coordinates $XY$.}
    \label{fig:eSq_1}
\end{figure*}

\begin{figure*}
    \centering
    \includegraphics[width=0.45\linewidth]{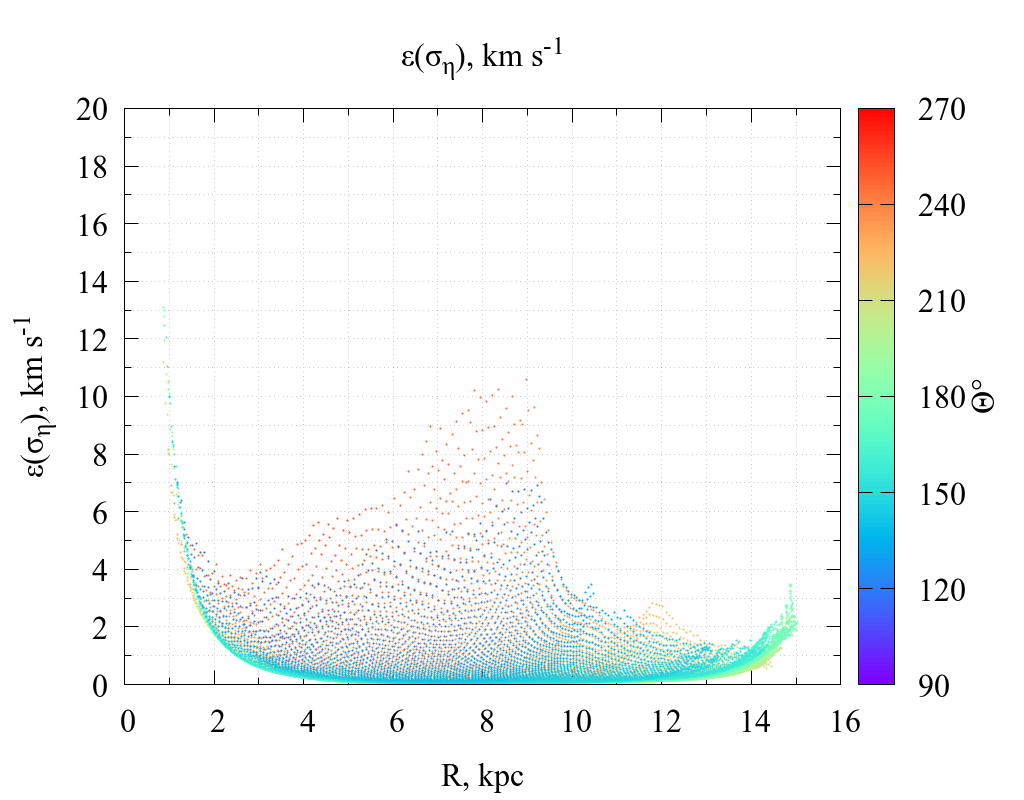}
    \includegraphics[width=0.45\linewidth]{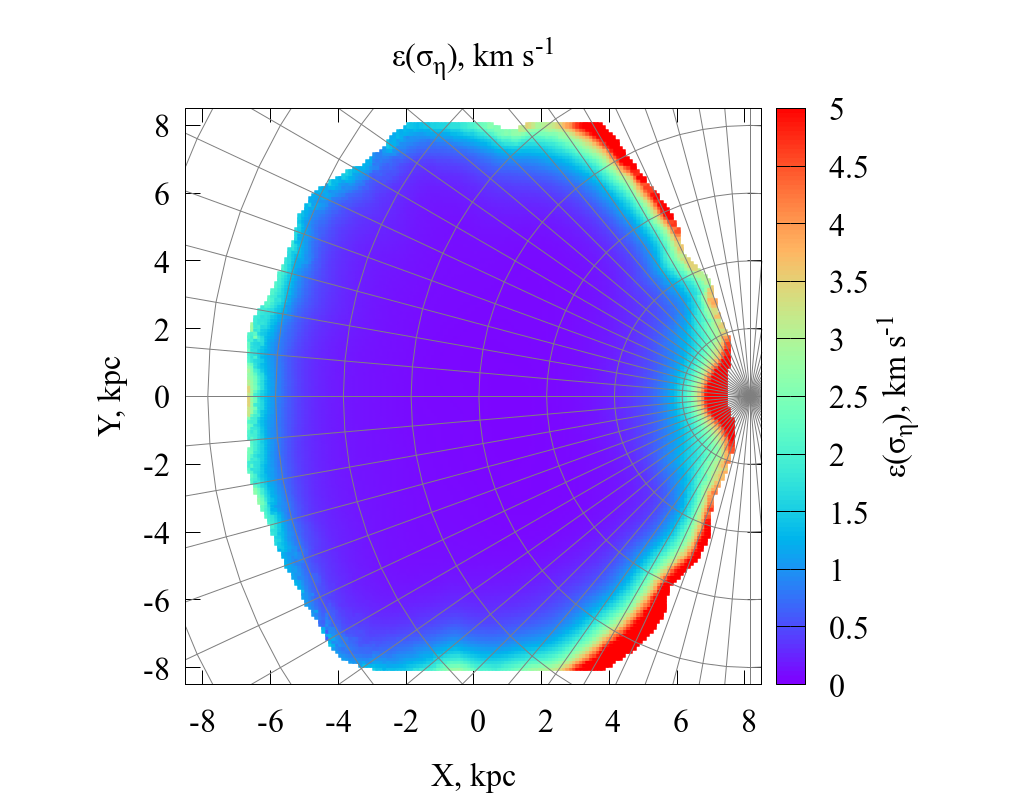}
    \caption{Left: uncertainties of the velocity ellipsoid semi-axis lengths $\varepsilon(\sigma_\eta)$ depending on the Galactocentric distance $R$, with $\theta$ shown in color. Right: the distribution map of $\varepsilon(\sigma_\eta)$ on the rectangular Galactocentric coordinates $XY$.}
    \label{fig:eSq_2}
\end{figure*}

\begin{figure*}
    \centering
    \includegraphics[width=0.45\linewidth]{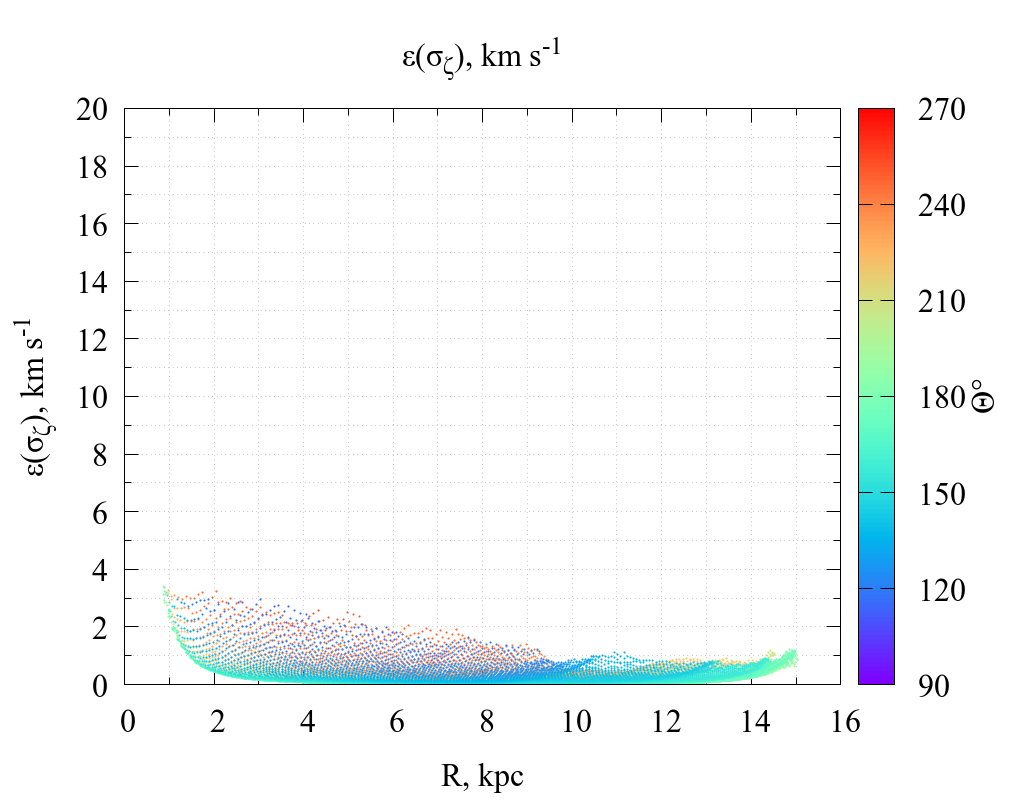}
    \includegraphics[width=0.45\linewidth]{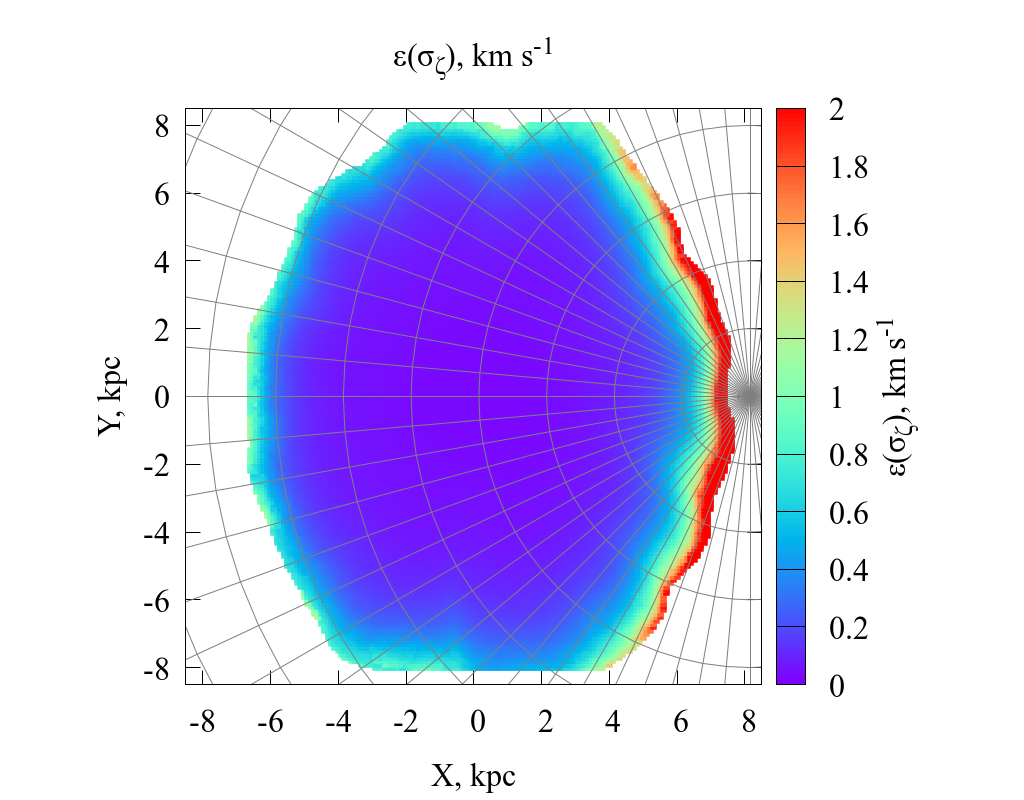}
    \caption{Left: uncertainties of the velocity ellipsoid semi-axis lengths $\varepsilon(\sigma_\zeta)$ depending on the Galactocentric distance $R$, with $\theta$ shown in color. Right: the distribution map of $\varepsilon(\sigma_\zeta)$ on the rectangular Galactocentric coordinates $XY$.}
    \label{fig:eSq_3}
\end{figure*}

\end{appendix}
\bsp 
\label{lastpage}
\end{document}